\documentclass[letter]{aa}  %

\usepackage{graphicx}
\usepackage[varg]{txfonts}
\usepackage{multirow}

\newcommand{\Ha}{H$\alpha$}
\newcommand{\Hb}{H$\beta$}

\begin{document}

   \title{A MUSE view of the asymmetric jet from HD 163296}


   \author{C.~Xie\inst{1,2}
          \and 
          S.~Y.~Haffert\inst{3, 2}\thanks{NASA Hubble Fellow} 
          \and
          J.~de~Boer\inst{2}
          \and
          M.~A.~Kenworthy\inst{2}
          \and
          J.~Brinchmann\inst{2,4}
          \and
          J.~Girard\inst{5}
          \and
          I.~A.~G.~Snellen\inst{2}
          \and
          C.~U.~Keller\inst{2}
         }
   \institute{Aix Marseille Univ, CNRS, CNES, LAM, Marseille, France 
            \email{chen.xie@lam.fr}
        \and
            Leiden Observatory, Leiden University, PO Box 9513, 2300 RA Leiden, The Netherlands
         \and
            Steward Observatory, 933 North Cherry Avenue, University of Arizona, Tucson, AZ 85721, USA
        \and
            Instituto de Astrof\'{i}sica e Ci\^{e}ncias do Espa\c{c}o, Universidade do Porto, CAUP, Rua das Estrelas, PT4150-762 Porto, Portugal
        \and 
            Space Telescope Science Institute, Baltimore 21218, MD, USA}

   \date{Received ---; accepted ---}

 
  \abstract
   {Jets and outflows are thought to play important roles in regulating star formation and disk evolution. An important question is how the jets are launched. HD~163296 is a well-studied Herbig Ae star that hosts proto-planet candidates, a protoplanetary disk, a protostellar jet, and a molecular outflow, which makes it an excellent laboratory for studying jets.
   }
   {We aim to characterize the jet at the inner regions and check if there are large differences with the features at large separations. A secondary objective is to demonstrate the performance of Multi Unit Spectroscopic Explorer (MUSE) in high-contrast imaging of extended line emission.}
   {MUSE in the narrow field mode (NFM) can provide observations at optical wavelengths with high spatial ($\sim$75 mas) and medium spectral ($R\sim$2500) resolution. With the high-resolution spectral differential imaging (HRSDI) technique, we can characterize the kinematic structures and physical conditions of jets down to 100 mas.}
   {We detect multiple atomic lines in two new knots, B3 and A4, at distances of \textless4\arcsec~from the host star with MUSE. 
   The derived $\dot{M}_{\rm jet} / \dot{M}_{\rm acc}$ is about 0.08 and 0.06 for knots B3 and A4, respectively. The observed [Ca~II]/[S~II] ratios indicate that there is no sign of dust grains at distances of \textless4\arcsec. Assuming the A4 knot traced the streamline, we can estimate a jet radius at the origin by fitting the half width half maximum (HWHM) of the jet, which sets an upper limit of 2.2 au on the size of the launching region. Although MUSE has the ability to detect the velocity shifts caused by high- and low-velocity components, we found no significant evidence of velocity decrease transverse to the jet direction in our 500~s MUSE observation.
   }
   {Our work demonstrates the capability of using MUSE NFM observations for the detailed study of stellar jets in the optical down to 100~mas. The derived $\dot{M}_{\rm jet} / \dot{M}_{\rm acc}$, no dust grain, and jet radius at the star support the magneto-centrifugal models as a launching mechanism for the jet.}


   \maketitle
%

\section{Introduction}

Protostellar jets have been observed in pre-main sequence stars across a wide range of masses \citep{Frank2014}, which have velocities of 50-400~km s$^{-1}$ and are collimated via the magnetic hoop stress by the magnetic field that is threading the disk \citep{Konigl2000, Ferreira2006, Ray2007}. The launching site of jets is thought to be at the innermost part of 
the disks within the 1~au scale \citep{Lee2017NatAs}, launched magneto-centrifugally from disks. Among the magneto-centrifugal models for jet launching, two competing models are the X-wind model \citep{Shu2000} and the disk 
wind model \citep{Konigl2000}. Adaptive-optics assisted integral-field spectrographs with medium spectral resolution can map the kinematics of jets in the inner region (less than a few hundred au), which can provide critical constraints in distinguishing between the X-wind and the disk wind models \citep{Frank2014}, as the ejection speed in the classical X-wind model is similar at all angles \citep{Shang2007}. In contrast with the X-wind model, the disk wind model can produce velocity decreases transverse to the jet direction \citep{Ferreira2006}. SINFONI observations of \object{DG\,Tau} show an onion-like velocity structure within 1\arcsec, disfavoring the classical X-wind model \citep{Agra-Amboage2011}. Here we focus on characterizing the jet in the inner region (\textless~400~au) of the Herbig Ae star \object{HD\,163296}.

HD~163296 is a 5-7 Myr old \citep{Montesinos2009, Vioque2018} Herbig Ae star, which is located at a distance of 101.5 $\pm$ 1.2~pc \citep{GaiaCollaboration2016, GaiaCollaboration2018}. Early observations by the Hubble Space Telescope 
discovered a bipolar jet, known as Herbig Haro (HH) object 409 \citep{Grady2000, Wassell2006} and a large dust disk with a size of $\sim$450 au \citep{Grady2000,  Wisniewski2008}. Since then, numerous works have studied the spatially resolved disk with ground-based telescopes  \citep{Fukagawa2010, Garufi2014, Guidi2016, Isella2016PhRvL, Garufi2017, Monnier2017, Muro-Arena2018, Guidi2018, DSHARP_IX_Isella2018, Rich2019}. In addition to the disk, the bipolar jet \object{HH\,409} shows complex structures with the size of a few thousand au. A series of HH knots were first detected in the point spread function (PSF) subtracted image \citep{Wassell2006}. \cite{Klaassen2013} found a double-corkscrew molecular outflow in CO 2-1 emission, which extends $\sim$1000 au along the flow direction. With VLT/X-shooter observations, \cite{Ellerbroek2014} presented the first optical to near-infrared spectra of the jet in HD~163296 down to 2\arcsec and detected more than 40 atomic lines. In this letter we present the spectral analysis of the jet from HD~163296 at <4\arcsec~with an unprecedented spatial resolution of $\sim$7~au, obtained with the optical integral-field spectrograph Multi Unit Spectroscopic Explorer (MUSE).

\section{Observations and data reduction}

HD~163296 was observed with 
MUSE \citep{Bacon2010SPIE} in narrow field mode (NFM) during the night of 
29 April~2019 (program 0103.C-0399(A), PI: de Boer). Laser tomography adaptive optics assisted NFM provided a spatial resolution of $\sim$75~mas with the spatial pixel (spaxel) size of 25~mas~$\times$~25~mas in a field of view of $7.5\arcsec \times 7.5\arcsec$, covering 480 nm - 930 nm with the spectral resolving power of 1740 at 480 nm and 3450 at 930 nm. The data were calibrated and reduced using the ESO MUSE pipeline, version 2.6. The details of the MUSE pipeline can be found in \cite{Weilbacher2020}. Three exposure settings were adopted to optimize the observations of this bright star with an {\sl R} band magnitude of 6.86 mag. The details of the MUSE observations can be found in Table~\ref{tab:obs_log}. Short exposures 
(2~s) were used to obtain the unsaturated stellar PSF and the stellar emission. Two long exposures of 250~s were obtained to perform the main analysis. Due to the brightness of the star, we masked the saturated data within a radius of 0.5\arcsec and replaced the data with the exposures of 8~s. We did not combine all the data with multiple exposure settings due to 
the read-out noise in 8s exposures, which degrades the signal-to-noise ratio (S/N) of the jet.

The jet emission is much fainter than the stellar halo. We used the high-resolution spectral differential imaging (HRSDI) technique to remove the stellar spectrum, which was first used in the discovery of the protoplanet \object{PDS\,70 c} \citep{Haffert2019}. HRSDI is suitable to retrieve sharp emission line features while removing the stellar halo.

\cite{Xie2020b}~found MUSE has instrumental issues that limit its performance for the application of high-contrast imaging. Such issues are more severe for objects with strong line emissions (i.e., strong \Ha~emission in HD~163296). We adopted a modified HRSDI that builds multiple reference spectra annularly (instead of single reference) with an extra wavelength calibration (see, \cite{Xie2020b}~and Appendix~\ref{appsec:modifiedHRSDI} 
for the details). 
Throughout the letter, the velocities are relative to the stellar rest frame, for which we adopt $v_{\rm LSR} = 5.8 \pm 0.2 $ km~s$^{-1}$ from \cite{Qi2011}. The dust extinction was considered to be zero along with the jet by \cite{Ellerbroek2014} and hence no extinction correction was applied to the data. A potential source of extinction is the protoplanetary disk, which is discussed in Sect.~\ref{subsec:contradict_to_16yr}.

\section{Analysis and characterization}


\label{sec:result}

\begin{figure*}
   \centering
    \includegraphics[width=0.98\textwidth]{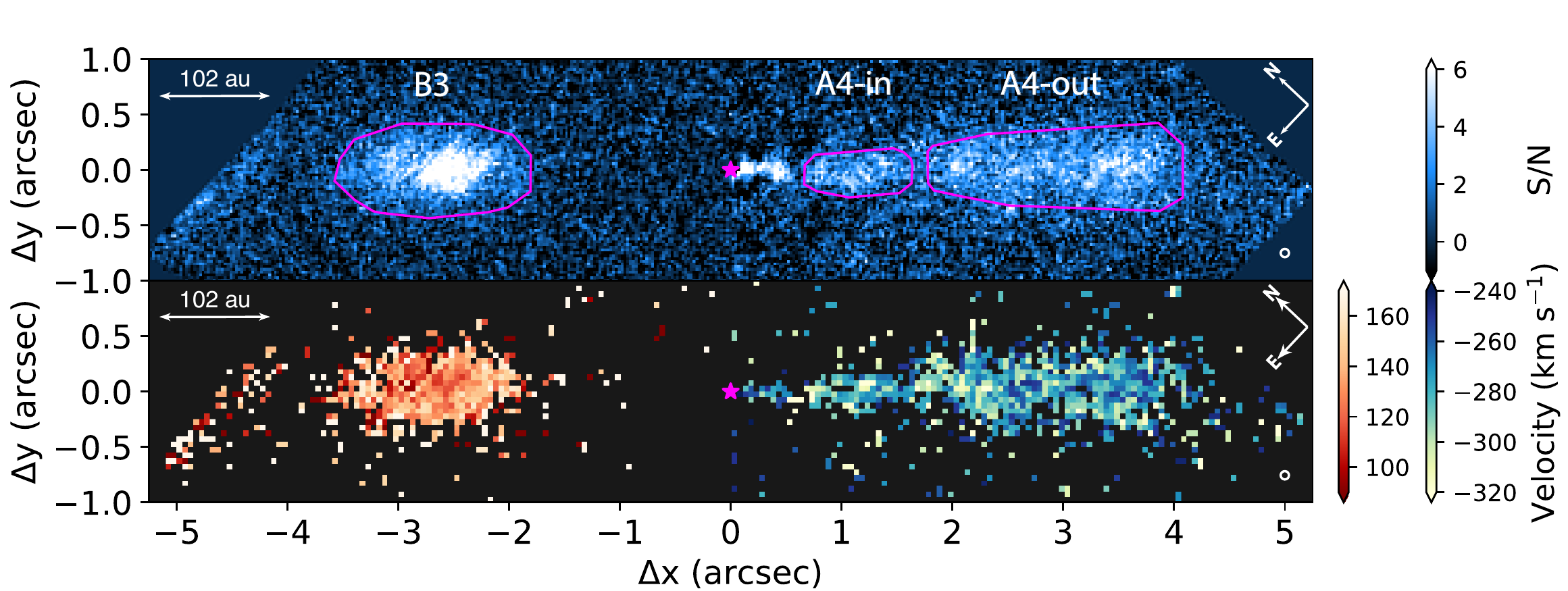}
      \caption{S/N (top) and radial velocity (bottom) maps of the asymmetric jet in HD 163296 detected by MUSE, traced by [S II] $\lambda$673 ($\Delta$x \textless~0) and \Ha~lines ($\Delta$x \textgreater~0), respectively. The red-shifted lobe is at $\Delta$x \textless~0 and the blue-shifted lobe is at $\Delta$x \textgreater~0. 
      The magenta polygon indicates the region where we measured the integrated jet flux.
      The spatial resolutions of the S/N and velocity maps are indicated by the white circles of $\sim$75~mas in diameter in the bottom right of each panel, which are $\sim$3 and $\sim$1.5 (2$\times$2 binning) spaxels across, respectively. Only spaxels that have the S/N of the line emission larger than 3$\sigma$ were used to construct the velocity map. The velocity uncertainty map is shown in Fig.\ref{velocity_uncertainty_map}. At the target distance of 101.5 pc, 1\arcsec corresponds to a scale of 102 au. The magenta star marks the star center. 
      }
         \label{jets_image}
\end{figure*}

\subsection{Detection of asymmetric jet at \textless 4\arcsec}

HD~163296 is one of the targets in a search for protoplanets with MUSE and as such was briefly discussed in \cite{Xie2020b}, showing only the morphology of the asymmetric jet. Fig.~\ref{jets_image} shows the S/N map of the asymmetric jet traced by the [S II] $\lambda$673 and \Ha~lines. It highlights the imaging capability of MUSE NFM that can probe the jet down to $\sim$100 mas after the removal of the stellar emission. The red-shifted and blue-shifted jet lobes are resolved in the northeast (NE) and southwest (SW) direction, respectively.

For knot B3, we find that the jet diameter is $\sim$0.4\arcsec. The jet diameter of knot A4 ranges from less than 0.075\arcsec to 0.65\arcsec. The jet diameter was estimated by performing one-dimensional Gaussian fits to the intensity image transverse to the jet direction, and then corrected for the effects of the PSF by subtracting the full width half maximum (FWHM)\footnote{The jet diameter $D_{\rm jet}^{2} = {\rm FWHM}_{\rm obs}^{2} - {\rm FWHM}_{\rm ins}^{2}$.}. The intensity image was binned in steps of 10 spaxels along the jet direction to increase the S/N. The knot A4 has a conical shape with a half opening angle of $\sim$5.1$^{\circ}$~at projected distances of \textless2.7\arcsec. An extrapolation to the origin of the jet suggests a launching radius of 0.95~$\pm$~1.27 au after correction of the 
inclination \citep[46.7$^{\circ}$;][]{DSHARP_IX_Isella2018}, and assuming 
these cones traced the streamlines (see, Fig.\ref{A4_in_1D}).  Due to the 
possible disk obscuration discussed in Sect.~\ref{subsec:contradict_to_16yr}, the half opening angle of the B3 knot is \textgreater 3.3$^{\circ}$~at projected distances of \textless2.5\arcsec.

Multiple atomic lines were detected and measured by fitting a single Gaussian in the spectral direction on the spatially integrated spectra, which are summarized in Table~\ref{tab:jet_line}. The corresponding line ratios 
are listed in Table~\ref{tab:ratio_and_conditions}. Fig.\ref{jet_spec} shows the integrated jet spectra where we masked the stellar \Ha~line ranging from 6561\AA - 6566\AA.

\begin{table}[th!]
\footnotesize
\caption{Line ratios and physical conditions.}             
\label{tab:ratio_and_conditions}      
\centering                          
\begin{tabular}{lccc}        
\hline\hline                 
Name & B3  & A4-in & A4-out   \\    
\hline
\multicolumn{4}{c}{Line ratios}\\
\hline                        
{\Ha / \Hb}                             &  --               & --              
& 2.66 $\pm$ 0.46 \\
{[S II]671 / [S II]673}             &  0.68 $\pm$ 0.05  & 0.53 $\pm$ 0.16 
& 0.80 $\pm$ 0.13 \\
{[O I]630 / \Ha}~                   &  --               & 0.33 $\pm$ 0.07 
& 0.12 $\pm$ 0.02 \\
{[S II]\tablefootmark{a} / \Ha}     &  --               & 0.62 $\pm$ 0.09 
& 0.80 $\pm$ 0.07 \\ 
{[N II]658 / [O I]630}              & 0.45 $\pm$ 0.05   & 1.95 $\pm$ 0.43 
& 3.72 $\pm$ 0.69 \\
{[S II]673 / \Ha}                   & --                & 0.40 $\pm$ 0.07 
& 0.46 $\pm$ 0.05 \\
{[N II]658 / [S II]\tablefootmark{a}} & 0.21 $\pm$ 0.02 & 1.04 $\pm$ 0.17 
& 0.56 $\pm$ 0.07 \\
{[Ca II]729 / [S II]673}            & 0.58 $\pm$ 0.03   & 0.36 $\pm$ 0.11 
& 0.37 $\pm$ 0.06 \\
\hline
\multicolumn{4}{c}{Physical conditions}\\
\hline
Position                      & $\sim$2.5\arcsec    & $\sim$1.1\arcsec  & 
$\sim$3\arcsec \\
$R_{\rm jet}$~(au)                 & 21.0 $\pm$ 0.6       & 14.4 $\pm$ 1.1      & 33.9 $\pm$  2.4 \\
$n_{\rm e}$ (cm$^{-3}$)       & 1715 $\pm$ 436                & 5131 $\pm$ 8586              & 978 $\pm$ 541 \\
$n_{\rm H,~pre}$ (cm$^{-3}$)  & 1000                & 1000              & 
100 \\
$B$ ($\mu$ G)                 & 10 - 100            & 100               & 
10 - 100 \\
$v_{\rm shock}$ (km~s$^{-1}$) & $\sim$50            & 80 - 100          & 
80 - 100 \\
$<I>$                         & 0.11 $\pm$ 0.01           & 0.42 $\pm$ 0.06         & 0.56 $\pm$ 0.02 \\
$<C>$                         & 23 $\pm$ 1            & 18 $\pm$ 2        
  & 18 $\pm$ 3 \\
$<n_{\rm H}>$ (cm$^{-3}$)               & 3251 $\pm$ 880                & 
2879 $\pm$ 4938              & 411 $\pm$ 230 \\
$\dot{M}$ (10$^{-10}$ M$_{\odot}$~yr$^{-1}$)    & 6.3 $\pm$ 1.7   & 5.8 $\pm$ 9.8           & 4.5 $\pm$ 2.6 \\

\hline
\end{tabular}
\tablefoot{The detailed derivation of the physical parameters can be found in the main text and in Appendix~\ref{appsec:phy_condition}.
\tablefoottext{a}{Here [S II] is the sum of [S II] $\lambda$671 and [S II] $\lambda$673 lines. }

}
\end{table}  

\subsection{Kinematics}
\label{sec:velocity_map}

The wavelength offset of the emission line was converted to a radial velocity based on the Doppler effect. Besides the different morphology of the jet, the radial velocities are also asymmetric, with the blue-shifted lobe being faster ($\sim$280 km~s$^{-1}$) and the red-shifted lobe being slower ($\sim$130 km~s$^{-1}$). The radial velocity in the blue-shifted lobe obtained with MUSE is consistent with the VLT/X-shooter result of 270~$\pm$~20 km~s$^{-1}$ in \cite{Ellerbroek2014} at larger angular separations (i.e., \textgreater~5\arcsec). However, the red-shifted lobe observed with MUSE has a lower radial velocity of around 130 km~s$^{-1}$, which is lower than 170~$\pm$~20 km~s$^{-1}$ that was obtained at distances of \textgreater~9\arcsec  \citep{Ellerbroek2014}. Nevertheless, our result is similar 
to the B2 knot\footnote{The definition of the jet knots can be found in \cite{Wassell2006} and \cite{Gunther2013}.} that was observed at a closer distance of 3.3\arcsec~in 2011 \citep{Gunther2013}, indicating a change in the radial velocity for the red-shifted lobe. 


Considering the large extent ($\sim$4.5\arcsec) of the blue-shifted lobe, 
we also measured the radial velocities for multiple lines as a function of the angular separation, which is shown in Fig.~\ref{A4_in_1D}. The velocities of \Ha~are consistent with other lines that were not affected by the strong stellar line emission, except for the radial velocity in \Ha~at 
$\sim$0.62\arcsec that shows a larger velocity. This is likely caused by the residual from the wing of the stellar \Ha~line emission, which broadens the jet \Ha~line toward shorter wavelengths.

We constructed the 2D velocity map of the [S II] $\lambda$673 and the \Ha~lines for the B3 and A4 knots, respectively, with a binning of 2 $\times$ 2 spaxels (see, Fig.~\ref{jets_image}). 
We excluded the binned spaxels that the S/N of the line emission is less than 3$\sigma$ and then performed a single Gaussian fitting in the spectral direction for each remaining spaxel. The velocity uncertainty map can be found in Appendix~\ref{velocity_uncertainty_map}, which shows the fitting 
error caused by the noise. Considering the uncertainty of around 10~km s$^{-1}$, the velocity of the A4 knot traced by \Ha~shows a relatively uniform distribution down to 100~mas. We note that only the \Ha~emission line has enough S/N to make the 2D velocity map for the A4 knot. As for the B3 knot, multiple lines (i.e., [S II] $\lambda$671, [N II] $\lambda$658, [O I] $\lambda$636, and [Ca II] $\lambda$729) have results consistent with [S II] $\lambda$673 in Fig.\ref{jets_image}, showing a uniform distribution without a significant velocity decrease transverse to the jet direction.

\subsection{Physical conditions}

\begin{figure}
   \centering
    \includegraphics[width=0.48\textwidth]{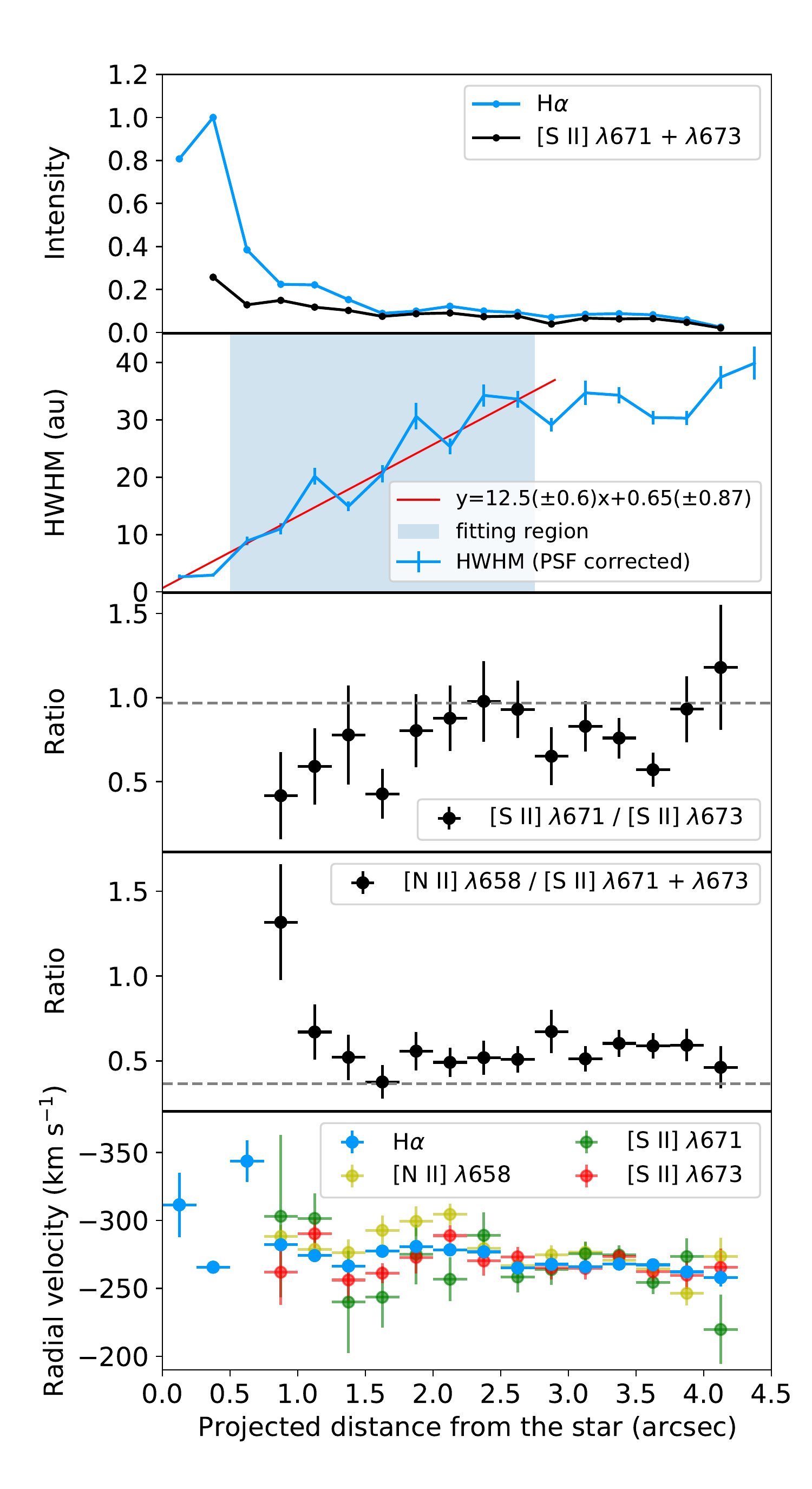}
      \caption{Variation in the line properties along the blue-shifted jet lobe. From top to bottom panel: Intensity profiles of the lines, PSF corrected jet half width half maximum (HWHM) at \Ha, observed ratios [S~II] 
$\lambda$671/[S~II] $\lambda$671 and [N~II] $\lambda$658/[S~II] ($\lambda$671+$\lambda$673), and radial velocities of lines. The red line shows the fitting result of the data in the fitting region, which gives a jet radius of 0.65~$\pm$~0.87 au at the star, before correcting for the inclination effect. The gray dashed lines represent the minimum [S~II] ratio and the maximum [N~II]/[S~II] ratio observed at the outer region in \cite{Ellerbroek2014}, respectively.
      }
         \label{A4_in_1D}
\end{figure}

Table~\ref{tab:ratio_and_conditions} shows the physical conditions for the B3 and A4 knots, derived from commonly used line ratios (see Appendix~\ref{appsec:phy_condition} for details of the estimation). The most used indicator for the electron density is the line ratio of [S~II] $\lambda$671 and [S~II] $\lambda$673 \citep{Proxauf2014}, assuming an electron temperature of 10 000~K \citep{Ellerbroek2014}. From the ratio of the [S II] lines, we obtained an electron density of about 1700 cm$^{-3}$ in knot B3. The electron density in knot A4 decreases along the jet lobe, from about 5100 cm$^{-3}$ in knot A4-in at $\sim$1.1\arcsec, to about 1000 cm$^{-3}$ 
in knot A4-out at $\sim$3\arcsec. At the larger distance (\textgreater4\arcsec), \cite{Ellerbroek2014} reported an electron density of $\sim$500~cm$^{-3}$ for both the red-shifted and blue-shifted lobes and they also found that the electron density decreases with the distance from the star. We also observed a decreasing trend at the inner region. To illustrate this decreasing trend in knot A4, we show the line ratio as a function of angular distance in Fig.~\ref{A4_in_1D}. The decreasing electron densities 
are revealed by an increasing [S~II] $\lambda$671/[S~II] $\lambda$673 ratio along the jet lobe. This confirms that the electron densities decreases 
along both jet lobes.

To estimate the mass-loss rate, we followed \cite{Hartigan1994}:
\begin{equation}
\label{equ_Mjet_A}
\dot{M}_{\rm jet} = \mu m_{\rm H}<n_{\rm H}>\pi R_{\rm J}^{2}v_{\rm J}.\end{equation}
Here $\mu$ = 1.24 is the mean molecular weight; $m_{\rm H}$, <$n_{\rm H}$>, $R_{\rm J}$, and $v_{\rm J}$ are the mass of the hydrogen atom, average density, the jet radius, and the jet velocity ($v_{\rm J} = |v_{\rm 
r}|/{\rm cos}i$), respectively. The average density can be estimated from 
electron density as $<n_{\rm H}> = <C>^{-1/2}<I>^{-1}n_{\rm e}$ with the ionization fraction <$I$> and mean compression factor <$C$> listed in Table~\ref{tab:ratio_and_conditions}. 
The derived $\dot{M}_{\rm jet}$ are $(6.3 \pm 1.7) \times 10^{-10}$ M$_{\odot}$~yr$^{-1}$, $(5.8 \pm 9.8)  \times 10^{-10}$ M$_{\odot}$~yr$^{-1}$, 
and $(4.5 \pm 2.6)  \times 10^{-10}$ M$_{\odot}$~yr$^{-1}$ for 
knots B3, A4-in, and A4-out, respectively. The mass-loss rate in the inner 
and outer part of the A4 knot are consistent with each other. The large uncertainty of $\dot{M}_{\rm jet}$ at the A4-in knot is caused by the relatively large uncertainty in the line flux of [S~II]~ $\lambda$671, which can be significantly improved with a longer integration time. And the mass-loss rate in knot B3 is a factor of 1.5 higher than knot A4. While the jet knots we measured are not the same ones as in \cite{Gunther2013} and \cite{Ellerbroek2014}, our measurements show similar mass-loss rates to previous knots at larger distances.

The mass accretion rate of the star can be estimated from the line width as 
\begin{equation}
\label{equ_Macc_line_width}
{\rm log}(\dot{M}_{\rm acc}) = -12.89(\pm0.3) + 9.7(\pm0.7) \times 10^{-3}~{\rm H}\alpha(10\%),
\end{equation}
where H$\alpha$(10\%) is the full width at 10\% maximum of the \Ha~line in km~s$^{-1}$ and the unit of $\dot{M}_{\rm acc}$ is M$_{\odot}$~yr$^{-1}$ \citep{Natta2004}. 
Based on the unsaturated short exposures, the intrinsic stellar \Ha~10\% width is 491 $\pm$ 23 km~s$^{-1}$ after deconvolving for the spectral line spread function (LSF)\footnote{H$\alpha$(10\%) = $\sqrt{ {\rm H}\alpha(10\%)_{\rm obs}^{2} -  {\rm H}\alpha(10\%)_{\rm ins}^{2}}$. The ${\rm H}\alpha(10\%)_{\rm ins}$ is 211 km~s$^{-1}$ at 6562.8\AA~\citep{Bacon2017}.}, which indicates a mass accretion rate of $7.5 \times 10^{-9.0 \pm 0.5}$ M$_{\odot}$~yr$^{-1}$. Assuming a constant $\dot{M}_{\rm acc}$, the $\dot{M}_{\rm jet} / \dot{M}_{\rm acc}$ ratios are about 0.08 $\pm$ 0.09 and 0.06 $\pm$ 0.08 for knots B3 and A4, respectively. Our measurements at 
the inner region (\textless 4\arcsec) are consistent with the $\dot{M}_{\rm jet} / \dot{M}_{\rm acc}$ ratios measured at the outer region (\textgreater 4\arcsec) by \cite{Ellerbroek2014}.

Shocks can destroy the dust grains and release the refractory atoms (such as Ca) back to the gas phase, which changes the [Ca II]/[S II] ratio. As a result, the line ratio of [Ca II] $\lambda$729 and [S II] $\lambda$673 traces the abundance of calcium in the gas phase and hence the presence of dust grains \citep{Podio2006}. We can look for the depletion of Ca by comparing the observed line ratio with a model prediction. As shown in Table~\ref{tab:ratio_and_conditions}, the measured [Ca II]/[S II] ratios are about 0.6 and 0.4 for the knots B3 and A4, respectively. These observed ratios are above the predicted line ratios (\textless~0.4) when the Ca$^{+}$ ionization fraction is equal to the hydrogen ionization (see, Fig.~4 in \cite{Ellerbroek2014}), indicating the absence of dust grains in the jet at \textless 4\arcsec. From the \Ha/\Hb~ratio of 2.7~$\pm$~0.5 in the A4-out knot at $\sim$3\arcsec, we can conclude that there is no significant amount of dust extinction since the ratio is close to the predicted value of $\sim$3 for the shock velocity of 80-100 km~s$^{-1}$ \citep{Hartigan1994}.

\section{Discussion}
\label{sec:discussion}

\subsection{A new knot possibly contradicting the 16 yr period}
\label{subsec:contradict_to_16yr}

Based on observations obtained before 2014, a period of $16.0~\pm~0.7$ yr 
and proper motions of 0.28\arcsec~yr$^{-1}$ (receding) and 0.49\arcsec~yr$^{-1}$ (approaching) were found in the jet knots that were regularly spaced \citep{Ellerbroek2014}. The knots in the receding jet lobe shows a roughly equal space interval of 4.5\arcsec. The B2 knot was detected at 3.4\arcsec~in 2013 \citep{Ellerbroek2014} and was expected to be located at $\sim$5.0\arcsec~by the time of our observation in 2019, assuming a constant proper motion of 0.28\arcsec~yr$^{-1}$. If the 16 yr period is real, the next knot B3 would be located at \textless~1\arcsec~in 2019. Interestingly, we detected a new knot B3 at $\sim$2.5\arcsec~in 2019, which is too close to the B2 knot and seems to contradict the 16 yr period. The clear drop in the flux on both sides of the jet axis suggests that we either detected a new knot or we only detected the brighter part of the B2 knot.

Three possible explanations can account for the presence of the new red-shifted knot B3 detected in this work. The first explanation is that the new knot we detect belongs to the previous knot B2. The second hypothesis is that only the outer part of the new knot was detected due to the disk obscuration and that the period has not changed. This is supported as the 
red-shifted knot is close to the edge of the dust disk around 193 au\footnote{The surface brightness of the dust disk has three distinct power-law 
trends \citep{Wisniewski2008}. A relatively constant intensity expends up 
to 2.8\arcsec, which corresponds to a projected separation of 1.9\arcsec or 193 au in the NE direction. Then the surface brightness follows an exponential $\sim r^{-4}$ decrease from $\sim$2.9\arcsec to $\sim$4.4\arcsec, following with termination of the detection at \textgreater4.5\arcsec.} 
\citep{Wassell2006, Wisniewski2008}. Similarly, disk obscuration of a receding jet lobe has been observed in DG~Tau \citep{Agra-Amboage2011, White2014}. Unfortunately, the \Ha~emission in B3 is contaminated by the residual of the stellar emission due to the instrumental issues, which limit our use of the \Ha/\Hb~ratio to characterize the extinction. The last explanation is an actual change in the period. During the launch of the B2 knot around 2002, an enhanced near-infrared excess was observed and thought to be caused by the launch of the dust cloud \citep{Ellerbroek2014}. A similar enhanced near-infrared excess was observed in the Ks-band in 2011 and 2012 \citep{Ellerbroek2014, Rich2020}, which maybe coincide with the launch of the newly detected B3 knot around 2010. Furthermore, no significant enhancement of a near-infrared excess was found between\emph{} 2016-2018 \citep{Rich2020}. Future observations with X-shooter will be able to determine whether the B3 knot is still part of the previous B2 knot. Higher S/N observations with MUSE NFM can also be used to determine the influence of the obscuration from the dust disk.


\subsection{Velocity uncertainty}
\label{subsec:velocity_unvertainty}

The velocity shifts transverse to the jet direction are on the order of about 50 km~s$^{-1}$ between high- and low-velocity components\footnote{Throughout the paper, we refer to a high-velocity component as its velocity 
\textgreater100 km~s$^{-1}$ and a low-velocity component as its velocity \textless100 km~s$^{-1}$.} and less than 15~km~s$^{-1}$ for jet rotations. The error on the velocity measurement consists of two components, the calibration errors in the wavelength solution and the centroid fitting error of the line center caused by the measurement noise. For the case of HD~163296, the fitting error is smaller than 10 km~s$^{-1}$ for spaxels with S/Ns larger 
than 5 (see Fig.~\ref{velocity_uncertainty_map}). The total integration time we used is only 500~s. With an integration time of 3.5~hr, the fitting 
error would be reduced by a factor of 5 as the fitting error $\sigma~\propto~(S/N)^{-1}$ \citep{Porter2004}, reaching the level of 2 km~s$^{-1}$. This is only valid if the error budget is dominated by the photon noise, which is the case for MUSE at separations of  \textless~1\arcsec~\citep{Xie2020b}.

The offset of the wavelength solution in MUSE can reach the level of 0.1\AA~\citep{Weilbacher2020}, which corresponds to $\sim$ 5~km~s$^{-1}$ at \Ha. For HD~163296, the average offset found during an extra round of wavelength calibration is around 0.25~\AA~and 0.35~\AA~for exposures of 250~s 
and 8~s, respectively. Tellurics (5577~\AA and 6300~\AA) were used in the 
MUSE pipeline to correct the offset caused by the temperature difference in the instrument between nighttime science images and daytime calibrations \citep{Weilbacher2020}. Without the additional calibration, the wavelength calibration errors are around 10 and 15 km~s$^{-1}$ at~\Ha~for exposures of 250~s and 8~s, respectively.  To further reduce the calibration error, we applied an extra wavelength calibration using the stellar \Ha~line, which is much stronger than telluric emission lines. Unfortunately, being a Herbig Ae star, HD~163296 does not have other prominent emission or absorption features. Hence, we cannot estimate the residual offset after our extra wavelength calibration.

Overall, MUSE can be used to observe the velocity shifts caused by high- and low-velocity components at \textless~1\arcsec. However, the relatively large offset of the wavelength solution prevents us from measuring any potential jet rotation at a 5$\sigma$ confidence level in the dataset of HD~163296. Future instruments will be able to measure the jet rotation in the visible and the near-infrared. For optical wavelengths, VLT/MAVIS \citep{mcdermid2020mavis} will provide the necessary spatial and spectral resolving power ($R\approx10000$), while VLT/ERIS \citep{davies2018eris}, which has a similar resolving power, can be used to probe emission lines in the near-infrared. If we look further into the future, HARMONI \citep{rodrigues2018harmoni}, the first generation IFU for the ELT, will provide even higher spatial and spectral resolution in the near-infrared.

\subsection{Better measurement of $\dot{M}_{\rm jet} / \dot{M}_{\rm acc}$ 
ratios with MUSE}

The mass-loss rate of the jet and the $\dot{M}_{\rm jet} / \dot{M}_{\rm acc}$ ratio are fundamental parameters to understand the jet efficiency in regulating the star formation and disk evolution \citep{Frank2014, Nisini2018}. In general, the measurements of the  $\dot{M}_{\rm jet} / \dot{M}_{\rm acc}$ ratio suffer from nonsimultaneous observations. There is a delay of several years to decades before jet knots are separated enough from their host star so that instruments such as X-shooter can observe the knots. MUSE can provide measurements of emission lines very close to a star (\textless~1\arcsec), which reduces the time interval between the measurements of the $\dot{M}_{\rm jet}$ and $ \dot{M}_{\rm acc}$, leading to a more accurate ratio. Furthermore, MUSE can also measure the jet radius, which is a key parameter in determining $\dot{M}_{\rm jet}$. It should be noted that the characterization of the innermost region depends on the S/N because MUSE does not have a coronagraph to block the stellar light.

\section{Conclusions}
\label{sec:conclusions}

We present the detailed spectral analysis of two new knots, B3 and A4, found by MUSE. Multiple atomic emission lines have been detected at \textless~4\arcsec after the removal of the stellar emission with HRSDI. With the MUSE NFM observation, we confirm that the HH~409 jet is asymmetric in morphology, velocity, and physical conditions down to 100~mas. The A4 knot is blue-shifted and has a radial velocity of about 280 km~s$^{-1}$, which is consistent with knots at larger distances. The B3 knot is red-shifted and has a lower radial velocity of about 130 km~s$^{-1}$. This knot also has a lower velocity compared to other receding knots at larger distances. Based on the increasing [S~II] $\lambda$671/[S~II] $\lambda$673 line ratios, we confirm that the electron densities decrease along both the receding and approaching jet lobes. The derived $\dot{M}_{\rm jet} / \dot{M}_{\rm 
acc}$ ratios are 0.08 and 0.06 for knots B3 and A4, respectively. No sign 
of dust grains is found in any of the knots based on the measured [Ca~II]/[S~II] ratios. Assuming the knot A4 traces the streamline, we find a jet 
launching radius of $1.0 \pm1.3$ au. The derived $\dot{M}_{\rm jet} / \dot{M}_{\rm acc}$ ratios, the absence of dust grains in the high-velocity jet, and the jet launching radius support the magneto-centrifugal models for jet launching. Although MUSE has the ability to detect the velocity shifts caused by high- and low-velocity components, we found no significant evidence of velocity decrease transverse to the jet direction.

Our work demonstrates the capability of using MUSE NFM observations for detailed studies of the morphology, kinematic structure, and physical conditions of stellar jets at optical wavelengths down to separations of 100 mas. Future observations of HD~163296 with MUSE and with longer integration times are recommended to get better measurements of the kinematic structure of the jet.

\begin{acknowledgements}
 
This work is based on observations collected at the European Organisation 
for Astronomical Research in the Southern Hemisphere under ESO programmes 
0103.C-0399(A). This research made use of Astropy\footnote{http://www.astropy.org}, a community-developed core Python package for Astronomy \citep{astropy:2013, astropy:2018}.

Support for this work was provided by NASA through the NASA Hubble Fellowship grant \#HST-HF2-51436.001-A awarded by the Space Telescope Science Institute, which is operated by the Association of Universities for Research in Astronomy, Incorporated, under NASA contract NAS5-26555.

\end{acknowledgements}

%
%

\bibliographystyle{aa}
\bibliography{msc2}

\begin{thebibliography}{48}
\expandafter\ifx\csname natexlab\endcsname\relax\def\natexlab#1{#1}\fi

\bibitem[{{Agra-Amboage} {et~al.}(2011){Agra-Amboage}, {Dougados}, {Cabrit}, \&
  {Reunanen}}]{Agra-Amboage2011}
{Agra-Amboage}, V., {Dougados}, C., {Cabrit}, S., \& {Reunanen}, J. 2011, \aap,
  532, A59

\bibitem[{{Astropy Collaboration} {et~al.}(2013){Astropy Collaboration},
  {Robitaille}, {Tollerud}, {Greenfield}, {Droettboom}, {Bray}, {Aldcroft},
  {Davis}, {Ginsburg}, {Price-Whelan}, {Kerzendorf}, {Conley}, {Crighton},
  {Barbary}, {Muna}, {Ferguson}, {Grollier}, {Parikh}, {Nair}, {Unther},
  {Deil}, {Woillez}, {Conseil}, {Kramer}, {Turner}, {Singer}, {Fox}, {Weaver},
  {Zabalza}, {Edwards}, {Azalee Bostroem}, {Burke}, {Casey}, {Crawford},
  {Dencheva}, {Ely}, {Jenness}, {Labrie}, {Lim}, {Pierfederici}, {Pontzen},
  {Ptak}, {Refsdal}, {Servillat}, \& {Streicher}}]{astropy:2013}
{Astropy Collaboration}, {Robitaille}, T.~P., {Tollerud}, E.~J., {et~al.} 2013,
  \aap, 558, A33

\bibitem[{{Bacon} {et~al.}(2010){Bacon}, {Accardo}, {Adjali}, {Anwand},
  {Bauer}, {Biswas}, {Blaizot}, {Boudon}, {Brau-Nogue}, {Brinchmann},
  {Caillier}, {Capoani}, {Carollo}, {Contini}, {Couderc}, {Daguis{\'e}},
  {Deiries}, {Delabre}, {Dreizler}, {Dubois}, {Dupieux}, {Dupuy}, {Emsellem},
  {Fechner}, {Fleischmann}, {Fran{\c c}ois}, {Gallou}, {Gharsa}, {Glindemann},
  {Gojak}, {Guiderdoni}, {Hansali}, {Hahn}, {Jarno}, {Kelz}, {Koehler},
  {Kosmalski}, {Laurent}, {Le Floch}, {Lilly}, {Lizon}, {Loupias}, {Manescau},
  {Monstein}, {Nicklas}, {Olaya}, {Pares}, {Pasquini}, {P{\'e}contal-Rousset},
  {Pell{\'o}}, {Petit}, {Popow}, {Reiss}, {Remillieux}, {Renault}, {Roth},
  {Rupprecht}, {Serre}, {Schaye}, {Soucail}, {Steinmetz}, {Streicher}, {Stuik},
  {Valentin}, {Vernet}, {Weilbacher}, {Wisotzki}, \& {Yerle}}]{Bacon2010SPIE}
{Bacon}, R., {Accardo}, M., {Adjali}, L., {et~al.} 2010, in procspie, Vol.
  7735, Ground-based and Airborne Instrumentation for Astronomy III, 773508

\bibitem[{{Bacon} {et~al.}(2017){Bacon}, {Conseil}, {Mary}, {Brinchmann},
  {Shepherd}, {Akhlaghi}, {Weilbacher}, {Piqueras}, {Wisotzki}, {Lagattuta},
  {Epinat}, {Guerou}, {Inami}, {Cantalupo}, {Courbot}, {Contini}, {Richard},
  {Maseda}, {Bouwens}, {Bouch{\'e}}, {Kollatschny}, {Schaye}, {Marino},
  {Pello}, {Herenz}, {Guiderdoni}, \& {Carollo}}]{Bacon2017}
{Bacon}, R., {Conseil}, S., {Mary}, D., {et~al.} 2017, \aap, 608, A1

\bibitem[{{Davies} {et~al.}(2018){Davies}, {Esposito}, {Schmid}, {Taylor},
  {Agapito}, {Agudo Berbel}, {Baruffolo}, {Biliotti}, {Biller}, {Black},
  {Boehle}, {Briguglio}, {Buron}, {Carbonaro}, {Cortes}, {Cresci},
  {Deysenroth}, {Di Cianno}, {Di Rico}, {Doelman}, {Dolci}, {Dorn},
  {Eisenhauer}, {Fantinel}, {Ferruzzi}, {Feuchtgruber}, {F{\"o}rster
  Schreiber}, {Gao}, {Gemperlein}, {Genzel}, {George}, {Gillessen}, {Giordano},
  {Glauser}, {Glindemann}, {Grani}, {Hartl}, {Heijmans}, {Henry}, {Huber},
  {Kasper}, {Keller}, {Kenworthy}, {K{\"u}hn}, {Kuntschner}, {Lightfoot},
  {Lunney}, {MacIntosh}, {Mannucci}, {March}, {Neeser}, {Patapis}, {Pearson},
  {Plattner}, {Puglisi}, {Quanz}, {Rau}, {Riccardi}, {Salasnich}, {Schubert},
  {Snik}, {Sturm}, {Valentini}, {Waring}, {Wiezorrek}, \&
  {Xompero}}]{davies2018eris}
{Davies}, R., {Esposito}, S., {Schmid}, H.~M., {et~al.} 2018, in Society of
  Photo-Optical Instrumentation Engineers (SPIE) Conference Series, Vol. 10702,
  Ground-based and Airborne Instrumentation for Astronomy VII, ed. C.~J.
  {Evans}, L.~{Simard}, \& H.~{Takami}, 1070209

\bibitem[{{Ellerbroek} {et~al.}(2014){Ellerbroek}, {Podio}, {Dougados},
  {Cabrit}, {Sitko}, {Sana}, {Kaper}, {de Koter}, {Klaassen}, {Mulders},
  {Mendigut{\'\i}a}, {Grady}, {Grankin}, {van Winckel}, {Bacciotti}, {Russell},
  {Lynch}, {Hammel}, {Beerman}, {Day}, {Huelsman}, {Werren}, {Henden}, \&
  {Grindlay}}]{Ellerbroek2014}
{Ellerbroek}, L.~E., {Podio}, L., {Dougados}, C., {et~al.} 2014, \aap, 563, A87

\bibitem[{{Ferreira} {et~al.}(2006){Ferreira}, {Dougados}, \&
  {Cabrit}}]{Ferreira2006}
{Ferreira}, J., {Dougados}, C., \& {Cabrit}, S. 2006, \aap, 453, 785

\bibitem[{{Frank} {et~al.}(2014){Frank}, {Ray}, {Cabrit}, {Hartigan}, {Arce},
  {Bacciotti}, {Bally}, {Benisty}, {Eisl{\"o}ffel}, {G{\"u}del}, {Lebedev},
  {Nisini}, \& {Raga}}]{Frank2014}
{Frank}, A., {Ray}, T.~P., {Cabrit}, S., {et~al.} 2014, in Protostars and
  Planets VI, ed. H.~{Beuther}, R.~S. {Klessen}, C.~P. {Dullemond}, \&
  T.~{Henning}, 451

\bibitem[{{Fukagawa} {et~al.}(2010){Fukagawa}, {Tamura}, {Itoh}, {Oasa},
  {Kudo}, {Hayashi}, {Kato}, {Ootsubo}, {Itoh}, {Shibai}, \&
  {Hayashi}}]{Fukagawa2010}
{Fukagawa}, M., {Tamura}, M., {Itoh}, Y., {et~al.} 2010, \pasj, 62, 347

\bibitem[{{Gaia Collaboration} {et~al.}(2018){Gaia Collaboration}, {Brown},
  {Vallenari}, {Prusti}, {de Bruijne}, {Babusiaux}, {Bailer-Jones}, {Biermann},
  {Evans}, {Eyer}, {Jansen}, {Jordi}, {Klioner}, {Lammers}, {Lindegren},
  {Luri}, {Mignard}, {Panem}, {Pourbaix}, {Randich}, {Sartoretti}, {Siddiqui},
  {Soubiran}, {van Leeuwen}, {Walton}, {Arenou}, {Bastian}, {Cropper},
  {Drimmel}, {Katz}, {Lattanzi}, {Bakker}, {Cacciari}, {Casta{\~n}eda},
  {Chaoul}, {Cheek}, {De Angeli}, {Fabricius}, {Guerra}, {Holl}, {Masana},
  {Messineo}, {Mowlavi}, {Nienartowicz}, {Panuzzo}, {Portell}, {Riello},
  {Seabroke}, {Tanga}, {Th{\'e}venin}, {Gracia-Abril}, {Comoretto},
  {Garcia-Reinaldos}, {Teyssier}, {Altmann}, {Andrae}, {Audard},
  {Bellas-Velidis}, {Benson}, {Berthier}, {Blomme}, {Burgess}, {Busso},
  {Carry}, {Cellino}, {Clementini}, {Clotet}, {Creevey}, {Davidson}, {De
  Ridder}, {Delchambre}, {Dell'Oro}, {Ducourant},
  {Fern{\'a}ndez-Hern{\'a}ndez}, {Fouesneau}, {Fr{\'e}mat}, {Galluccio},
  {Garc{\'\i}a-Torres}, {Gonz{\'a}lez-N{\'u}{\~n}ez}, {Gonz{\'a}lez-Vidal},
  {Gosset}, {Guy}, {Halbwachs}, {Hambly}, {Harrison}, {Hern{\'a}ndez},
  {Hestroffer}, {Hodgkin}, {Hutton}, {Jasniewicz}, {Jean-Antoine-Piccolo},
  {Jordan}, {Korn}, {Krone-Martins}, {Lanzafame}, {Lebzelter}, {L{\"o}ffler},
  {Manteiga}, {Marrese}, {Mart{\'\i}n-Fleitas}, {Moitinho}, {Mora}, {Muinonen},
  {Osinde}, {Pancino}, {Pauwels}, {Petit}, {Recio-Blanco}, {Richards},
  {Rimoldini}, {Robin}, {Sarro}, {Siopis}, {Smith}, {Sozzetti}, {S{\"u}veges},
  {Torra}, {van Reeven}, {Abbas}, {Abreu Aramburu}, {Accart}, {Aerts},
  {Altavilla}, {{\'A}lvarez}, {Alvarez}, {Alves}, {Anderson}, {Andrei},
  {Anglada Varela}, {Antiche}, {Antoja}, {Arcay}, {Astraatmadja}, {Bach},
  {Baker}, {Balaguer-N{\'u}{\~n}ez}, {Balm}, {Barache}, {Barata}, {Barbato},
  {Barblan}, {Barklem}, {Barrado}, {Barros}, {Barstow}, {Bartholom{\'e}
  Mu{\~n}oz}, {Bassilana}, {Becciani}, {Bellazzini}, {Berihuete}, {Bertone},
  {Bianchi}, {Bienaym{\'e}}, {Blanco-Cuaresma}, {Boch}, {Boeche}, {Bombrun},
  {Borrachero}, {Bossini}, {Bouquillon}, {Bourda}, {Bragaglia}, {Bramante},
  {Breddels}, {Bressan}, {Brouillet}, {Br{\"u}semeister}, {Brugaletta},
  {Bucciarelli}, {Burlacu}, {Busonero}, {Butkevich}, {Buzzi}, {Caffau},
  {Cancelliere}, {Cannizzaro}, {Cantat-Gaudin}, {Carballo}, {Carlucci},
  {Carrasco}, {Casamiquela}, {Castellani}, {Castro-Ginard}, {Charlot},
  {Chemin}, {Chiavassa}, {Cocozza}, {Costigan}, {Cowell}, {Crifo}, {Crosta},
  {Crowley}, {Cuypers}, {Dafonte}, {Damerdji}, {Dapergolas}, {David}, {David},
  {de Laverny}, {De Luise}, {De March}, {de Martino}, {de Souza}, {de Torres},
  {Debosscher}, {del Pozo}, {Delbo}, {Delgado}, {Delgado}, {Di Matteo},
  {Diakite}, {Diener}, {Distefano}, {Dolding}, {Drazinos}, {Dur{\'a}n},
  {Edvardsson}, {Enke}, {Eriksson}, {Esquej}, {Eynard Bontemps}, {Fabre},
  {Fabrizio}, {Faigler}, {Falc{\~a}o}, {Farr{\`a}s Casas}, {Federici},
  {Fedorets}, {Fernique}, {Figueras}, {Filippi}, {Findeisen}, {Fonti},
  {Fraile}, {Fraser}, {Fr{\'e}zouls}, {Gai}, {Galleti}, {Garabato},
  {Garc{\'\i}a-Sedano}, {Garofalo}, {Garralda}, {Gavel}, {Gavras}, {Gerssen},
  {Geyer}, {Giacobbe}, {Gilmore}, {Girona}, {Giuffrida}, {Glass}, {Gomes},
  {Granvik}, {Gueguen}, {Guerrier}, {Guiraud}, {Guti{\'e}rrez-S{\'a}nchez},
  {Haigron}, {Hatzidimitriou}, {Hauser}, {Haywood}, {Heiter}, {Helmi}, {Heu},
  {Hilger}, {Hobbs}, {Hofmann}, {Holland}, {Huckle}, {Hypki}, {Icardi},
  {Jan{\ss}en}, {Jevardat de Fombelle}, {Jonker}, {Juh{\'a}sz}, {Julbe},
  {Karampelas}, {Kewley}, {Klar}, {Kochoska}, {Kohley}, {Kolenberg},
  {Kontizas}, {Kontizas}, {Koposov}, {Kordopatis}, {Kostrzewa-Rutkowska},
  {Koubsky}, {Lambert}, {Lanza}, {Lasne}, {Lavigne}, {Le Fustec}, {Le
  Poncin-Lafitte}, {Lebreton}, {Leccia}, {Leclerc}, {Lecoeur-Taibi},
  {Lenhardt}, {Leroux}, {Liao}, {Licata}, {Lindstr{\o}m}, {Lister}, {Livanou},
  {Lobel}, {L{\'o}pez}, {Managau}, {Mann}, {Mantelet}, {Marchal}, {Marchant},
  {Marconi}, {Marinoni}, {Marschalk{\'o}}, {Marshall}, {Martino}, {Marton},
  {Mary}, {Massari}, {Matijevi{\v{c}}}, {Mazeh}, {McMillan}, {Messina},
  {Michalik}, {Millar}, {Molina}, {Molinaro}, {Moln{\'a}r}, {Montegriffo},
  {Mor}, {Morbidelli}, {Morel}, {Morris}, {Mulone}, {Muraveva}, {Musella},
  {Nelemans}, {Nicastro}, {Noval}, {O'Mullane}, {Ord{\'e}novic},
  {Ord{\'o}{\~n}ez-Blanco}, {Osborne}, {Pagani}, {Pagano}, {Pailler},
  {Palacin}, {Palaversa}, {Panahi}, {Pawlak}, {Piersimoni}, {Pineau}, {Plachy},
  {Plum}, {Poggio}, {Poujoulet}, {Pr{\v{s}}a}, {Pulone}, {Racero}, {Ragaini},
  {Rambaux}, {Ramos-Lerate}, {Regibo}, {Reyl{\'e}}, {Riclet}, {Ripepi}, {Riva},
  {Rivard}, {Rixon}, {Roegiers}, {Roelens}, {Romero-G{\'o}mez}, {Rowell},
  {Royer}, {Ruiz-Dern}, {Sadowski}, {Sagrist{\`a} Sell{\'e}s}, {Sahlmann},
  {Salgado}, {Salguero}, {Sanna}, {Santana-Ros}, {Sarasso}, {Savietto},
  {Schultheis}, {Sciacca}, {Segol}, {Segovia}, {S{\'e}gransan}, {Shih},
  {Siltala}, {Silva}, {Smart}, {Smith}, {Solano}, {Solitro}, {Sordo}, {Soria
  Nieto}, {Souchay}, {Spagna}, {Spoto}, {Stampa}, {Steele},
  {Steidelm{\"u}ller}, {Stephenson}, {Stoev}, {Suess}, {Surdej}, {Szabados},
  {Szegedi-Elek}, {Tapiador}, {Taris}, {Tauran}, {Taylor}, {Teixeira},
  {Terrett}, {Teyssand ier}, {Thuillot}, {Titarenko}, {Torra Clotet}, {Turon},
  {Ulla}, {Utrilla}, {Uzzi}, {Vaillant}, {Valentini}, {Valette}, {van Elteren},
  {Van Hemelryck}, {van Leeuwen}, {Vaschetto}, {Vecchiato}, {Veljanoski},
  {Viala}, {Vicente}, {Vogt}, {von Essen}, {Voss}, {Votruba}, {Voutsinas},
  {Walmsley}, {Weiler}, {Wertz}, {Wevers}, {Wyrzykowski}, {Yoldas},
  {{\v{Z}}erjal}, {Ziaeepour}, {Zorec}, {Zschocke}, {Zucker}, {Zurbach}, \&
  {Zwitter}}]{GaiaCollaboration2018}
{Gaia Collaboration}, {Brown}, A.~G.~A., {Vallenari}, A., {et~al.} 2018, \aap,
  616, A1

\bibitem[{{Gaia Collaboration} {et~al.}(2016){Gaia Collaboration}, {Prusti},
  {de Bruijne}, {Brown}, {Vallenari}, {Babusiaux}, {Bailer-Jones}, {Bastian},
  {Biermann}, {Evans}, {Eyer}, {Jansen}, {Jordi}, {Klioner}, {Lammers},
  {Lindegren}, {Luri}, {Mignard}, {Milligan}, {Panem}, {Poinsignon},
  {Pourbaix}, {Randich}, {Sarri}, {Sartoretti}, {Siddiqui}, {Soubiran},
  {Valette}, {van Leeuwen}, {Walton}, {Aerts}, {Arenou}, {Cropper}, {Drimmel},
  {H{\o}g}, {Katz}, {Lattanzi}, {O'Mullane}, {Grebel}, {Holland}, {Huc},
  {Passot}, {Bramante}, {Cacciari}, {Casta{\~n}eda}, {Chaoul}, {Cheek}, {De
  Angeli}, {Fabricius}, {Guerra}, {Hern{\'a}ndez}, {Jean-Antoine-Piccolo},
  {Masana}, {Messineo}, {Mowlavi}, {Nienartowicz}, {Ord{\'o}{\~n}ez-Blanco},
  {Panuzzo}, {Portell}, {Richards}, {Riello}, {Seabroke}, {Tanga},
  {Th{\'e}venin}, {Torra}, {Els}, {Gracia-Abril}, {Comoretto},
  {Garcia-Reinaldos}, {Lock}, {Mercier}, {Altmann}, {Andrae}, {Astraatmadja},
  {Bellas-Velidis}, {Benson}, {Berthier}, {Blomme}, {Busso}, {Carry},
  {Cellino}, {Clementini}, {Cowell}, {Creevey}, {Cuypers}, {Davidson}, {De
  Ridder}, {de Torres}, {Delchambre}, {Dell'Oro}, {Ducourant}, {Fr{\'e}mat},
  {Garc{\'\i}a-Torres}, {Gosset}, {Halbwachs}, {Hambly}, {Harrison}, {Hauser},
  {Hestroffer}, {Hodgkin}, {Huckle}, {Hutton}, {Jasniewicz}, {Jordan},
  {Kontizas}, {Korn}, {Lanzafame}, {Manteiga}, {Moitinho}, {Muinonen},
  {Osinde}, {Pancino}, {Pauwels}, {Petit}, {Recio-Blanco}, {Robin}, {Sarro},
  {Siopis}, {Smith}, {Smith}, {Sozzetti}, {Thuillot}, {van Reeven}, {Viala},
  {Abbas}, {Abreu Aramburu}, {Accart}, {Aguado}, {Allan}, {Allasia},
  {Altavilla}, {{\'A}lvarez}, {Alves}, {Anderson}, {Andrei}, {Anglada Varela},
  {Antiche}, {Antoja}, {Ant{\'o}n}, {Arcay}, {Atzei}, {Ayache}, {Bach},
  {Baker}, {Balaguer-N{\'u}{\~n}ez}, {Barache}, {Barata}, {Barbier}, {Barblan},
  {Baroni}, {Barrado y Navascu{\'e}s}, {Barros}, {Barstow}, {Becciani},
  {Bellazzini}, {Bellei}, {Bello Garc{\'\i}a}, {Belokurov}, {Bendjoya},
  {Berihuete}, {Bianchi}, {Bienaym{\'e}}, {Billebaud}, {Blagorodnova},
  {Blanco-Cuaresma}, {Boch}, {Bombrun}, {Borrachero}, {Bouquillon}, {Bourda},
  {Bouy}, {Bragaglia}, {Breddels}, {Brouillet}, {Br{\"u}semeister},
  {Bucciarelli}, {Budnik}, {Burgess}, {Burgon}, {Burlacu}, {Busonero}, {Buzzi},
  {Caffau}, {Cambras}, {Campbell}, {Cancelliere}, {Cantat-Gaudin}, {Carlucci},
  {Carrasco}, {Castellani}, {Charlot}, {Charnas}, {Charvet}, {Chassat},
  {Chiavassa}, {Clotet}, {Cocozza}, {Collins}, {Collins}, {Costigan}, {Crifo},
  {Cross}, {Crosta}, {Crowley}, {Dafonte}, {Damerdji}, {Dapergolas}, {David},
  {David}, {De Cat}, {de Felice}, {de Laverny}, {De Luise}, {De March}, {de
  Martino}, {de Souza}, {Debosscher}, {del Pozo}, {Delbo}, {Delgado},
  {Delgado}, {di Marco}, {Di Matteo}, {Diakite}, {Distefano}, {Dolding}, {Dos
  Anjos}, {Drazinos}, {Dur{\'a}n}, {Dzigan}, {Ecale}, {Edvardsson}, {Enke},
  {Erdmann}, {Escolar}, {Espina}, {Evans}, {Eynard Bontemps}, {Fabre},
  {Fabrizio}, {Faigler}, {Falc{\~a}o}, {Farr{\`a}s Casas}, {Faye}, {Federici},
  {Fedorets}, {Fern{\'a}ndez-Hern{\'a}ndez}, {Fernique}, {Fienga}, {Figueras},
  {Filippi}, {Findeisen}, {Fonti}, {Fouesneau}, {Fraile}, {Fraser}, {Fuchs},
  {Furnell}, {Gai}, {Galleti}, {Galluccio}, {Garabato}, {Garc{\'\i}a-Sedano},
  {Gar{\'e}}, {Garofalo}, {Garralda}, {Gavras}, {Gerssen}, {Geyer}, {Gilmore},
  {Girona}, {Giuffrida}, {Gomes}, {Gonz{\'a}lez-Marcos},
  {Gonz{\'a}lez-N{\'u}{\~n}ez}, {Gonz{\'a}lez-Vidal}, {Granvik}, {Guerrier},
  {Guillout}, {Guiraud}, {G{\'u}rpide}, {Guti{\'e}rrez-S{\'a}nchez}, {Guy},
  {Haigron}, {Hatzidimitriou}, {Haywood}, {Heiter}, {Helmi}, {Hobbs},
  {Hofmann}, {Holl}, {Holland }, {Hunt}, {Hypki}, {Icardi}, {Irwin}, {Jevardat
  de Fombelle}, {Jofr{\'e}}, {Jonker}, {Jorissen}, {Julbe}, {Karampelas},
  {Kochoska}, {Kohley}, {Kolenberg}, {Kontizas}, {Koposov}, {Kordopatis},
  {Koubsky}, {Kowalczyk}, {Krone-Martins}, {Kudryashova}, {Kull}, {Bachchan},
  {Lacoste-Seris}, {Lanza}, {Lavigne}, {Le Poncin-Lafitte}, {Lebreton},
  {Lebzelter}, {Leccia}, {Leclerc}, {Lecoeur-Taibi}, {Lemaitre}, {Lenhardt},
  {Leroux}, {Liao}, {Licata}, {Lindstr{\o}m}, {Lister}, {Livanou}, {Lobel},
  {L{\"o}ffler}, {L{\'o}pez}, {Lopez-Lozano}, {Lorenz}, {Loureiro},
  {MacDonald}, {Magalh{\~a}es Fernandes}, {Managau}, {Mann}, {Mantelet},
  {Marchal}, {Marchant}, {Marconi}, {Marie}, {Marinoni}, {Marrese},
  {Marschalk{\'o}}, {Marshall}, {Mart{\'\i}n-Fleitas}, {Martino}, {Mary},
  {Matijevi{\v{c}}}, {Mazeh}, {McMillan}, {Messina}, {Mestre}, {Michalik},
  {Millar}, {Miranda}, {Molina}, {Molinaro}, {Molinaro}, {Moln{\'a}r},
  {Moniez}, {Montegriffo}, {Monteiro}, {Mor}, {Mora}, {Morbidelli}, {Morel},
  {Morgenthaler}, {Morley}, {Morris}, {Mulone}, {Muraveva}, {Musella},
  {Narbonne}, {Nelemans}, {Nicastro}, {Noval}, {Ord{\'e}novic},
  {Ordieres-Mer{\'e}}, {Osborne}, {Pagani}, {Pagano}, {Pailler}, {Palacin},
  {Palaversa}, {Parsons}, {Paulsen}, {Pecoraro}, {Pedrosa}, {Pentik{\"a}inen},
  {Pereira}, {Pichon}, {Piersimoni}, {Pineau}, {Plachy}, {Plum}, {Poujoulet},
  {Pr{\v{s}}a}, {Pulone}, {Ragaini}, {Rago}, {Rambaux}, {Ramos-Lerate},
  {Ranalli}, {Rauw}, {Read}, {Regibo}, {Renk}, {Reyl{\'e}}, {Ribeiro},
  {Rimoldini}, {Ripepi}, {Riva}, {Rixon}, {Roelens}, {Romero-G{\'o}mez},
  {Rowell}, {Royer}, {Rudolph}, {Ruiz-Dern}, {Sadowski}, {Sagrist{\`a}
  Sell{\'e}s}, {Sahlmann}, {Salgado}, {Salguero}, {Sarasso}, {Savietto},
  {Schnorhk}, {Schultheis}, {Sciacca}, {Segol}, {Segovia}, {Segransan},
  {Serpell}, {Shih}, {Smareglia}, {Smart}, {Smith}, {Solano}, {Solitro},
  {Sordo}, {Soria Nieto}, {Souchay}, {Spagna}, {Spoto}, {Stampa}, {Steele},
  {Steidelm{\"u}ller}, {Stephenson}, {Stoev}, {Suess}, {S{\"u}veges}, {Surdej},
  {Szabados}, {Szegedi-Elek}, {Tapiador}, {Taris}, {Tauran}, {Taylor},
  {Teixeira}, {Terrett}, {Tingley}, {Trager}, {Turon}, {Ulla}, {Utrilla},
  {Valentini}, {van Elteren}, {Van Hemelryck}, {van Leeuwen}, {Varadi},
  {Vecchiato}, {Veljanoski}, {Via}, {Vicente}, {Vogt}, {Voss}, {Votruba},
  {Voutsinas}, {Walmsley}, {Weiler}, {Weingrill}, {Werner}, {Wevers},
  {Whitehead}, {Wyrzykowski}, {Yoldas}, {{\v{Z}}erjal}, {Zucker}, {Zurbach},
  {Zwitter}, {Alecu}, {Allen}, {Allende Prieto}, {Amorim},
  {Anglada-Escud{\'e}}, {Arsenijevic}, {Azaz}, {Balm}, {Beck}, {Bernstein},
  {Bigot}, {Bijaoui}, {Blasco}, {Bonfigli}, {Bono}, {Boudreault}, {Bressan},
  {Brown}, {Brunet}, {Bunclark}, {Buonanno}, {Butkevich}, {Carret}, {Carrion},
  {Chemin}, {Ch{\'e}reau}, {Corcione}, {Darmigny}, {de Boer}, {de Teodoro}, {de
  Zeeuw}, {Delle Luche}, {Domingues}, {Dubath}, {Fodor}, {Fr{\'e}zouls},
  {Fries}, {Fustes}, {Fyfe}, {Gallardo}, {Gallegos}, {Gardiol}, {Gebran},
  {Gomboc}, {G{\'o}mez}, {Grux}, {Gueguen}, {Heyrovsky}, {Hoar}, {Iannicola},
  {Isasi Parache}, {Janotto}, {Joliet}, {Jonckheere}, {Keil}, {Kim},
  {Klagyivik}, {Klar}, {Knude}, {Kochukhov}, {Kolka}, {Kos}, {Kutka}, {Lainey},
  {LeBouquin}, {Liu}, {Loreggia}, {Makarov}, {Marseille}, {Martayan},
  {Martinez-Rubi}, {Massart}, {Meynadier}, {Mignot}, {Munari}, {Nguyen},
  {Nordlander}, {Ocvirk}, {O'Flaherty}, {Olias Sanz}, {Ortiz}, {Osorio},
  {Oszkiewicz}, {Ouzounis}, {Palmer}, {Park}, {Pasquato}, {Peltzer}, {Peralta},
  {P{\'e}turaud}, {Pieniluoma}, {Pigozzi}, {Poels}, {Prat}, {Prod'homme},
  {Raison}, {Rebordao}, {Risquez}, {Rocca-Volmerange}, {Rosen}, {Ruiz-Fuertes},
  {Russo}, {Sembay}, {Serraller Vizcaino}, {Short}, {Siebert}, {Silva},
  {Sinachopoulos}, {Slezak}, {Soffel}, {Sosnowska}, {Strai{\v{z}}ys}, {ter
  Linden}, {Terrell}, {Theil}, {Tiede}, {Troisi}, {Tsalmantza}, {Tur},
  {Vaccari}, {Vachier}, {Valles}, {Van Hamme}, {Veltz}, {Virtanen}, {Wallut},
  {Wichmann}, {Wilkinson}, {Ziaeepour}, \& {Zschocke}}]{GaiaCollaboration2016}
{Gaia Collaboration}, {Prusti}, T., {de Bruijne}, J.~H.~J., {et~al.} 2016,
  \aap, 595, A1

\bibitem[{{Garufi} {et~al.}(2017){Garufi}, {Meeus}, {Benisty}, {Quanz},
  {Banzatti}, {Kama}, {Canovas}, {Eiroa}, {Schmid}, {Stolker}, {Pohl},
  {Rigliaco}, {M{\'e}nard}, {Meyer}, {van Boekel}, \& {Dominik}}]{Garufi2017}
{Garufi}, A., {Meeus}, G., {Benisty}, M., {et~al.} 2017, \aap, 603, A21

\bibitem[{{Garufi} {et~al.}(2014){Garufi}, {Quanz}, {Schmid}, {Avenhaus},
  {Buenzli}, \& {Wolf}}]{Garufi2014}
{Garufi}, A., {Quanz}, S.~P., {Schmid}, H.~M., {et~al.} 2014, \aap, 568, A40

\bibitem[{{Grady} {et~al.}(2000){Grady}, {Devine}, {Woodgate}, {Kimble},
  {Bruhweiler}, {Boggess}, {Linsky}, {Plait}, {Clampin}, \&
  {Kalas}}]{Grady2000}
{Grady}, C.~A., {Devine}, D., {Woodgate}, B., {et~al.} 2000, \apj, 544, 895

\bibitem[{{Guidi} {et~al.}(2018){Guidi}, {Ruane}, {Williams}, {Mawet}, {Testi},
  {Zurlo}, {Absil}, {Bottom}, {Choquet}, {Christiaens}, {Femen{\'\i}a
  Castell{\'a}}, {Huby}, {Isella}, {Kastner}, {Meshkat}, {Reggiani}, {Riggs},
  {Serabyn}, \& {Wallack}}]{Guidi2018}
{Guidi}, G., {Ruane}, G., {Williams}, J.~P., {et~al.} 2018, \mnras, 479, 1505

\bibitem[{{Guidi} {et~al.}(2016){Guidi}, {Tazzari}, {Testi}, {de
  Gregorio-Monsalvo}, {Chandler}, {P{\'e}rez}, {Isella}, {Natta}, {Ortolani},
  {Henning}, {Corder}, {Linz}, {Andrews}, {Wilner}, {Ricci}, {Carpenter},
  {Sargent}, {Mundy}, {Storm}, {Calvet}, {Dullemond}, {Greaves}, {Lazio},
  {Deller}, \& {Kwon}}]{Guidi2016}
{Guidi}, G., {Tazzari}, M., {Testi}, L., {et~al.} 2016, \aap, 588, A112

\bibitem[{{G{\"u}nther} {et~al.}(2013){G{\"u}nther}, {Schneider}, \&
  {Li}}]{Gunther2013}
{G{\"u}nther}, H.~M., {Schneider}, P.~C., \& {Li}, Z.~Y. 2013, \aap, 552, A142

\bibitem[{{Haffert} {et~al.}(2019){Haffert}, {Bohn}, {de Boer}, {Snellen},
  {Brinchmann}, {Girard}, {Keller}, \& {Bacon}}]{Haffert2019}
{Haffert}, S.~Y., {Bohn}, A.~J., {de Boer}, J., {et~al.} 2019, Nature
  Astronomy, 3, 749

\bibitem[{{Hartigan} {et~al.}(1994){Hartigan}, {Morse}, \&
  {Raymond}}]{Hartigan1994}
{Hartigan}, P., {Morse}, J.~A., \& {Raymond}, J. 1994, \apj, 436, 125

\bibitem[{{Isella} {et~al.}(2016){Isella}, {Guidi}, {Testi}, {Liu}, {Li}, {Li},
  {Weaver}, {Boehler}, {Carperter}, {De Gregorio-Monsalvo}, {Manara}, {Natta},
  {P{\'e}rez}, {Ricci}, {Sargent}, {Tazzari}, \& {Turner}}]{Isella2016PhRvL}
{Isella}, A., {Guidi}, G., {Testi}, L., {et~al.} 2016, \prl, 117, 251101

\bibitem[{{Isella} {et~al.}(2018){Isella}, {Huang}, {Andrews}, {Dullemond},
  {Birnstiel}, {Zhang}, {Zhu}, {Guzm{\'a}n}, {P{\'e}rez}, {Bai}, {Benisty},
  {Carpenter}, {Ricci}, \& {Wilner}}]{DSHARP_IX_Isella2018}
{Isella}, A., {Huang}, J., {Andrews}, S.~M., {et~al.} 2018, \apjl, 869, L49

\bibitem[{{Klaassen} {et~al.}(2013){Klaassen}, {Juhasz}, {Mathews}, {Mottram},
  {De Gregorio-Monsalvo}, {van Dishoeck}, {Takahashi}, {Akiyama}, {Chapillon},
  {Espada}, {Hales}, {Hogerheijde}, {Rawlings}, {Schmalzl}, \&
  {Testi}}]{Klaassen2013}
{Klaassen}, P.~D., {Juhasz}, A., {Mathews}, G.~S., {et~al.} 2013, \aap, 555,
  A73

\bibitem[{{Konigl} \& {Pudritz}(2000)}]{Konigl2000}
{Konigl}, A. \& {Pudritz}, R.~E. 2000, in Protostars and Planets IV, ed.
  V.~{Mannings}, A.~P. {Boss}, \& S.~S. {Russell}, 759

\bibitem[{Kramida {et~al.}(2019)Kramida, {Yu.~Ralchenko}, Reader, \& {and NIST
  ASD Team}}]{NIST_ASD}
Kramida, A., {Yu.~Ralchenko}, Reader, J., \& {and NIST ASD Team}. 2019, {NIST
  Atomic Spectra Database (ver. 5.7.1), [Online]. Available:
  {\tt{https://physics.nist.gov/asd}} [2020, July 29]. National Institute of
  Standards and Technology, Gaithersburg, MD.}

\bibitem[{{Lee} {et~al.}(2017){Lee}, {Ho}, {Li}, {Hirano}, {Zhang}, \&
  {Shang}}]{Lee2017NatAs}
{Lee}, C.-F., {Ho}, P. T.~P., {Li}, Z.-Y., {et~al.} 2017, Nature Astronomy, 1,
  0152

\bibitem[{{McDermid} {et~al.}(2020){McDermid}, {Cresci}, {Rigaut}, {Bouret},
  {De Silva}, {Gullieuszik}, {Magrini}, {Mendel}, {Antoniucci}, {Bono},
  {Kamath}, {Monty}, {Baumgardt}, {Cortese}, {Fisher}, {Mannucci},
  {Migliorini}, {Sweet}, {Vanzella}, {Zibetti}, \& ~}]{mcdermid2020mavis}
{McDermid}, R.~M., {Cresci}, G., {Rigaut}, F., {et~al.} 2020, arXiv e-prints,
  arXiv:2009.09242

\bibitem[{{Monnier} {et~al.}(2017){Monnier}, {Harries}, {Aarnio}, {Adams},
  {Andrews}, {Calvet}, {Espaillat}, {Hartmann}, {Hinkley}, {Kraus}, {McClure},
  {Oppenheimer}, {Perrin}, \& {Wilner}}]{Monnier2017}
{Monnier}, J.~D., {Harries}, T.~J., {Aarnio}, A., {et~al.} 2017, \apj, 838, 20

\bibitem[{{Montesinos} {et~al.}(2009){Montesinos}, {Eiroa}, {Mora}, \&
  {Mer{\'\i}n}}]{Montesinos2009}
{Montesinos}, B., {Eiroa}, C., {Mora}, A., \& {Mer{\'\i}n}, B. 2009, \aap, 495,
  901

\bibitem[{{Muro-Arena} {et~al.}(2018){Muro-Arena}, {Dominik}, {Waters}, {Min},
  {Klarmann}, {Ginski}, {Isella}, {Benisty}, {Pohl}, {Garufi}, {Hagelberg},
  {Langlois}, {Menard}, {Pinte}, {Sezestre}, {van der Plas}, {Villenave},
  {Delboulb{\'e}}, {Magnard}, {M{\"o}ller-Nilsson}, {Pragt}, {Rabou}, \&
  {Roelfsema}}]{Muro-Arena2018}
{Muro-Arena}, G.~A., {Dominik}, C., {Waters}, L.~B.~F.~M., {et~al.} 2018, \aap,
  614, A24

\bibitem[{{Natta} {et~al.}(2004){Natta}, {Testi}, {Muzerolle}, {Randich},
  {Comer{\'o}n}, \& {Persi}}]{Natta2004}
{Natta}, A., {Testi}, L., {Muzerolle}, J., {et~al.} 2004, \aap, 424, 603

\bibitem[{{Nisini} {et~al.}(2018){Nisini}, {Antoniucci}, {Alcal{\'a}},
  {Giannini}, {Manara}, {Natta}, {Fedele}, \& {Biazzo}}]{Nisini2018}
{Nisini}, B., {Antoniucci}, S., {Alcal{\'a}}, J.~M., {et~al.} 2018, \aap, 609,
  A87

\bibitem[{{Podio} {et~al.}(2006){Podio}, {Bacciotti}, {Nisini},
  {Eisl{\"o}ffel}, {Massi}, {Giannini}, \& {Ray}}]{Podio2006}
{Podio}, L., {Bacciotti}, F., {Nisini}, B., {et~al.} 2006, \aap, 456, 189

\bibitem[{{Porter} {et~al.}(2004){Porter}, {Oudmaijer}, \&
  {Baines}}]{Porter2004}
{Porter}, J.~M., {Oudmaijer}, R.~D., \& {Baines}, D. 2004, \aap, 428, 327

\bibitem[{{Price-Whelan} {et~al.}(2018){Price-Whelan}, {Sip{\H{o}}cz},
  {G{\"u}nther}, {Lim}, {Crawford}, {Conseil}, {Shupe}, {Craig}, {Dencheva},
  {Ginsburg}, {VanderPlas}, {Bradley}, {P{\'e}rez-Su{\'a}rez}, {de Val-Borro},
  {Paper Contributors}, {Aldcroft}, {Cruz}, {Robitaille}, {Tollerud},
  {Coordination Committee}, {Ardelean}, {Babej}, {Bach}, {Bachetti}, {Bakanov},
  {Bamford}, {Barentsen}, {Barmby}, {Baumbach}, {Berry}, {Biscani}, {Boquien},
  {Bostroem}, {Bouma}, {Brammer}, {Bray}, {Breytenbach}, {Buddelmeijer},
  {Burke}, {Calderone}, {Cano Rodr{\'\i}guez}, {Cara}, {Cardoso}, {Cheedella},
  {Copin}, {Corrales}, {Crichton}, {D{\textquoteright}Avella}, {Deil},
  {Depagne}, {Dietrich}, {Donath}, {Droettboom}, {Earl}, {Erben}, {Fabbro},
  {Ferreira}, {Finethy}, {Fox}, {Garrison}, {Gibbons}, {Goldstein}, {Gommers},
  {Greco}, {Greenfield}, {Groener}, {Grollier}, {Hagen}, {Hirst}, {Homeier},
  {Horton}, {Hosseinzadeh}, {Hu}, {Hunkeler}, {Ivezi{\'c}}, {Jain}, {Jenness},
  {Kanarek}, {Kendrew}, {Kern}, {Kerzendorf}, {Khvalko}, {King}, {Kirkby},
  {Kulkarni}, {Kumar}, {Lee}, {Lenz}, {Littlefair}, {Ma}, {Macleod},
  {Mastropietro}, {McCully}, {Montagnac}, {Morris}, {Mueller}, {Mumford},
  {Muna}, {Murphy}, {Nelson}, {Nguyen}, {Ninan}, {N{\"o}the}, {Ogaz}, {Oh},
  {Parejko}, {Parley}, {Pascual}, {Patil}, {Patil}, {Plunkett}, {Prochaska},
  {Rastogi}, {Reddy Janga}, {Sabater}, {Sakurikar}, {Seifert}, {Sherbert},
  {Sherwood-Taylor}, {Shih}, {Sick}, {Silbiger}, {Singanamalla}, {Singer},
  {Sladen}, {Sooley}, {Sornarajah}, {Streicher}, {Teuben}, {Thomas},
  {Tremblay}, {Turner}, {Terr{\'o}n}, {van Kerkwijk}, {de la Vega}, {Watkins},
  {Weaver}, {Whitmore}, {Woillez}, {Zabalza}, \& {Contributors}}]{astropy:2018}
{Price-Whelan}, A.~M., {Sip{\H{o}}cz}, B.~M., {G{\"u}nther}, H.~M., {et~al.}
  2018, \aj, 156, 123

\bibitem[{{Proxauf} {et~al.}(2014){Proxauf}, {{\"O}ttl}, \&
  {Kimeswenger}}]{Proxauf2014}
{Proxauf}, B., {{\"O}ttl}, S., \& {Kimeswenger}, S. 2014, \aap, 561, A10

\bibitem[{{Qi} {et~al.}(2011){Qi}, {D'Alessio}, {{\"O}berg}, {Wilner},
  {Hughes}, {Andrews}, \& {Ayala}}]{Qi2011}
{Qi}, C., {D'Alessio}, P., {{\"O}berg}, K.~I., {et~al.} 2011, \apj, 740, 84

\bibitem[{{Ray} {et~al.}(2007){Ray}, {Dougados}, {Bacciotti}, {Eisl{\"o}ffel},
  \& {Chrysostomou}}]{Ray2007}
{Ray}, T., {Dougados}, C., {Bacciotti}, F., {Eisl{\"o}ffel}, J., \&
  {Chrysostomou}, A. 2007, in Protostars and Planets V, ed. B.~{Reipurth},
  D.~{Jewitt}, \& K.~{Keil}, 231

\bibitem[{{Rich} {et~al.}(2019){Rich}, {Wisniewski}, {Currie}, {Fukagawa},
  {Grady}, {Sitko}, {Pikhartova}, {Hashimoto}, {Abe}, {Brand ner}, {Brandt},
  {Carson}, {Chilcote}, {Dong}, {Feldt}, {Goto}, {Groff}, {Guyon}, {Hayano},
  {Hayashi}, {Hayashi}, {Henning}, {Hodapp}, {Ishii}, {Iye}, {Janson},
  {Jovanovic}, {Kand ori}, {Kasdin}, {Knapp}, {Kudo}, {Kusakabe}, {Kuzuhara},
  {Kwon}, {Lozi}, {Martinache}, {Matsuo}, {Mayama}, {McElwain}, {Miyama},
  {Morino}, {Moro-Martin}, {Nakagawa}, {Nishimura}, {Pyo}, {Serabyn}, {Suto},
  {Russel}, {Suzuki}, {Takami}, {Takato}, {Terada}, {Thalmann}, {Turner},
  {Uyama}, {Wagner}, {Watanabe}, {Yamada}, {Takami}, {Usuda}, \&
  {Tamura}}]{Rich2019}
{Rich}, E.~A., {Wisniewski}, J.~P., {Currie}, T., {et~al.} 2019, \apj, 875, 38

\bibitem[{{Rich} {et~al.}(2020){Rich}, {Wisniewski}, {Sitko}, {Grady}, {Tobin},
  \& {Fukagawa}}]{Rich2020}
{Rich}, E.~A., {Wisniewski}, J.~P., {Sitko}, M.~L., {et~al.} 2020, \apj, 902, 4

\bibitem[{Rodrigues {et~al.}(2018)Rodrigues, Capone, Earle, Foster, Hidalgo,
  Lewis, Lynn, O'brien, Tosh, George, Accardo, Alvarez, Conzelmann, Hopgood,
  Clarke, Schnetler, Tecza, \& Thatte}]{rodrigues2018harmoni}
Rodrigues, M., Capone, J., Earle, A., {et~al.} 2018, in Ground-based and
  Airborne Instrumentation for Astronomy VII, ed. C.~J. Evans, L.~Simard, \&
  H.~Takami, Vol. 10702, International Society for Optics and Photonics (SPIE),
  2991 -- 3000

\bibitem[{{Shang} {et~al.}(2007){Shang}, {Li}, \& {Hirano}}]{Shang2007}
{Shang}, H., {Li}, Z.~Y., \& {Hirano}, N. 2007, in Protostars and Planets V,
  ed. B.~{Reipurth}, D.~{Jewitt}, \& K.~{Keil}, 261

\bibitem[{{Shu} {et~al.}(2000){Shu}, {Najita}, {Shang}, \& {Li}}]{Shu2000}
{Shu}, F.~H., {Najita}, J.~R., {Shang}, H., \& {Li}, Z.~Y. 2000, in Protostars
  and Planets IV, ed. V.~{Mannings}, A.~P. {Boss}, \& S.~S. {Russell}, 789--814

\bibitem[{{Vioque} {et~al.}(2018){Vioque}, {Oudmaijer}, {Baines},
  {Mendigut{\'\i}a}, \& {P{\'e}rez-Mart{\'\i}nez}}]{Vioque2018}
{Vioque}, M., {Oudmaijer}, R.~D., {Baines}, D., {Mendigut{\'\i}a}, I., \&
  {P{\'e}rez-Mart{\'\i}nez}, R. 2018, \aap, 620, A128

\bibitem[{{Wassell} {et~al.}(2006){Wassell}, {Grady}, {Woodgate}, {Kimble}, \&
  {Bruhweiler}}]{Wassell2006}
{Wassell}, E.~J., {Grady}, C.~A., {Woodgate}, B., {Kimble}, R.~A., \&
  {Bruhweiler}, F.~C. 2006, \apj, 650, 985

\bibitem[{{Weilbacher} {et~al.}(2020){Weilbacher}, {Palsa}, {Streicher},
  {Bacon}, {Urrutia}, {Wisotzki}, {Conseil}, {Husemann}, {Jarno}, {Kelz},
  {P{\'e}contal-Rousset}, {Richard}, {Roth}, {Selman}, \&
  {Vernet}}]{Weilbacher2020}
{Weilbacher}, P.~M., {Palsa}, R., {Streicher}, O., {et~al.} 2020, \aap, 641,
  A28

\bibitem[{{White} {et~al.}(2014){White}, {Bicknell}, {McGregor}, \&
  {Salmeron}}]{White2014}
{White}, M.~C., {Bicknell}, G.~V., {McGregor}, P.~J., \& {Salmeron}, R. 2014,
  \mnras, 442, 28

\bibitem[{{Wisniewski} {et~al.}(2008){Wisniewski}, {Clampin}, {Grady},
  {Ardila}, {Ford}, {Golimowski}, {Illingworth}, \& {Krist}}]{Wisniewski2008}
{Wisniewski}, J.~P., {Clampin}, M., {Grady}, C.~A., {et~al.} 2008, \apj, 682,
  548

\bibitem[{{Xie} {et~al.}(2020){Xie}, {Haffert}, {de Boer}, {Kenworthy},
  {Brinchmann}, {Girard}, {Snellen}, \& {Keller}}]{Xie2020b}
{Xie}, C., {Haffert}, S.~Y., {de Boer}, J., {et~al.} 2020, \aap, 644, A149

\end{thebibliography}

\begin{appendix}

\section{The observations}

\begin{table}[th!]
\caption{MUSE NFM observations of HD~163296.}             
\label{tab:obs_log}      
\centering                          
\begin{tabular}{ c c c c c c }        
\hline\hline                 
 $t_{\rm DIT} \times n_{\rm DIT}$$^{a}$ & On-source time & Derotation & Frames per rot.  \\    
    & (s) & (degree) & \\
\hline                        
 2 $\times$ 4 &  8 & 0 &  4  \\
 8 $\times$ 40 &  320 & 45 &  8  \\
 250 $\times$ 2 &  500 & 0 &  2  \\

\hline
\end{tabular}
\tablefoot{$^{(a)}$~$t_{\rm DIT}$ is the exposure time per image frame in the 
unit of seconds and $n_{\rm DIT}$ is the number of image frames.
}
\end{table}

\section{Stellar emission subtraction}
\label{appsec:modifiedHRSDI}

The key to obtaining a higher contrast ratio is to subtract the observed stellar emission as accurately as possible with the best-matched reference spectra. Due to the instrumental issues of MUSE, the observed stellar line-to-continuum ratio varies across the field if stars have strong line emission, which is the case for HD~163296. We follow the modified HRSDI method described in \cite{Xie2020b}, which can effectively correct for such issues. Modified HRSDI only adjusts the way of building the reference spectrum, compared with standard HRSDI described in \cite{Haffert2019}. According to the detailed description in \cite{Xie2020b}, we built reference spectra for several annuli. However, before the reference spectra were estimated, we applied an additional wavelength calibration using the strong stellar emission line (i.e., \Ha). The regions of the jet were masked when we built the stellar reference spectra to avoid influencing the stellar spectra with the jet emission. To avoid over-subtraction of the jet emission, no principal component analysis (PCA) was used for the correction of the high-order effect, while the residual remains relatively small thanks to the better-matched reference spectra and additional wavelength calibration. 

For the blue part of the spectrum (4772~\AA~- 5779~\AA), the only prominent feature in the stellar spectrum is a strong and broad absorption at \Hb. Such a broad feature, which is five times wider than \Ha~emission line, is not suitable for determining the offset of the wavelength solution. Therefore, we did not perform the extra wavelength calibration for the blue part of the spectrum.

As is shown in \cite{Xie2020b}, the instrumental issues led to strips in the images after removing the stellar emission. In the case of 250~s data without derotation (see, Table~\ref{tab:obs_log}), such strips are prominent spatially in the five horizontal IFUs in which the stellar PSF center is located and existed in the spectral channels around \Ha~(6550~\AA~- 6577~\AA) and \Hb~(4856.61~\AA~- 4864.11~\AA). To further remove these strips, we applied a median filter spatially in each row among the five IFUs with a length of 55 spaxels. To avoid the influence of the jet emission, the jet was masked not only spatially but also spectrally because the jet is only present at certain wavelengths with a 10\%~width of $\sim$6~\AA.

\begin{figure*}
   \centering
    \includegraphics[width=0.98\textwidth]{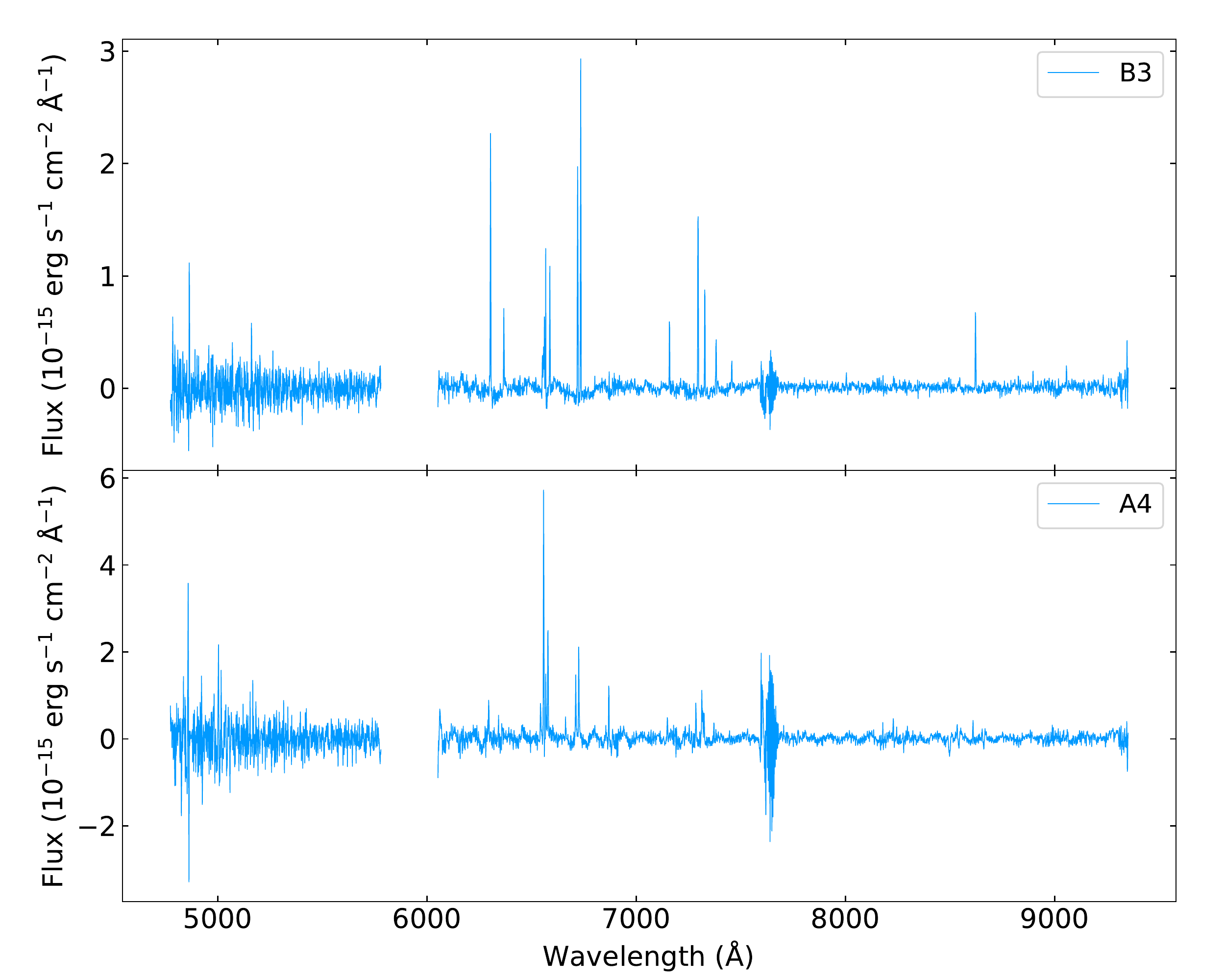}
      \caption{MUSE integrated spectra of the northeast (B3, top panel) and southwest (A4, bottom panel) jet knots in HD~163296, covering blue (4800 -- 5750~\AA) and red (6080 -- 9300~\AA) parts of spectra. The wavelength between 5780 \AA -- 6050~\AA~ was blocked to avoid contamination from the laser guide stars. The wavelength between 6561~\AA-- 6566~\AA~ was masked because of instrumental issues. 
      The region where we measured the integrated jet flux is shown in Fig.~\ref{jets_image}. The line properties are presented in Tables~\ref{tab:ratio_and_conditions} and~\ref{tab:jet_line}. 
      }
         \label{jet_spec}
\end{figure*}

Due to the poor AO performance at shorter wavelengths, the residual noise at the blue part of the spectrum is much larger than that at longer wavelengths. In addition to the noise, we also noticed a low-frequency variation in the residual with a period on the order of $\sim$60~\AA~in the spectral direction. Hence, we adopted a high-pass filter only for the blue part of the integrated jet spectra, which are presented in Fig.\ref{jet_spec}.

\section{Observed emission lines}

\begin{table*}[th!]
\caption{Jet properties.}             
\label{tab:jet_line}      
\centering                          
\begin{tabular}{l c c c c c c c}        
\hline\hline                 
name & line & $\lambda_{\textrm{rest frame}}$ & $\lambda_{\textrm{obs}}$ & radial velocity & line flux\tablefoottext{a} & FWHM\tablefootmark{b}   \\    
 &  & (\AA) & (\AA) & (km~s$^{-1}$) & 1 $\times$ 10$^{-16} $ (erg~s$^{-1}$~cm$^{-2}$) & (km~s$^{-1}$) \\
\hline
\multirow{14}{*}{\textbf{B3}}   
& \Hb  &        4861.35 &       4863.42   &   121.6 $\pm$ 7.4   &   32.5 $\pm$ 3.9  
 &   92.4 $\pm$ 60.4     \\
& --  &         -- &    5160.38   &   --   &   22.3 $\pm$ 5.0   &   149.7 $\pm$ 37.8       \\
& [O I]  &      6300.304  &     6303.11   &   127.4 $\pm$ 2.0   &   82.9 $\pm$ 2.6   &   125.5 $\pm$ 4.6  \\
& [O I] &       6363.776 &      6366.55   &   124.8 $\pm$ 6.6   &   24.7 $\pm$ 2.7 
  &   123.8 $\pm$ 15.0   \\
& [N II]  &     6583.45  &      6586.44   &   130.1 $\pm$ 5.8   &   36.9 $\pm$ 3.8   &   113.4 $\pm$ 13.1         \\
& [S II]  &     6716.440 &      6719.41   &   126.8 $\pm$ 3.1   &   70.9 $\pm$ 3.9   &   113.7 $\pm$ 7.2  \\
& [S II]  &     6730.816  &     6733.86   &   129.6 $\pm$ 2.1   &   104.2 $\pm$ 3.9   &   114.7 $\pm$ 5.0  \\
& [Fe II] &     7155.1742 &     7158.39   &   129.0 $\pm$ 5.1   &   21.5 $\pm$ 2.4   &   96.6 $\pm$ 13.7  \\
& [Ca II] &     7291.47  &      7294.72   &   127.9 $\pm$ 2.4   &   60.5 $\pm$ 2.8   &   105.1 $\pm$ 5.9  \\
& [Ca II] &     7323.89  &      7327.13   &   126.9 $\pm$ 4.0   &   32.9 $\pm$ 2.8   &   98.8 $\pm$ 10.2  \\
&  -- &         -- &    7381.07   &   --   &   17.3 $\pm$ 2.4   &   113.3 $\pm$ 18.5       \\
&  [Fe II] &    7452.54 &       7456.03   &   134.6 $\pm$ 19.3   &   10.3 $\pm$ 3.3   &   124.8 $\pm$ 45.7         \\
&  [Fe II] &    8616.9498  &    8620.89   &   131.4 $\pm$ 1.8   &   25.9 $\pm$ 1.1   &   87.7 $\pm$ 4.6   \\
\hline
\multirow{11}{*}{\textbf{A4-in}}
& \Hb\tablefootmark{c}  &       4861.35 &       4857.8   &   -224.9 $\pm$ 13.7   & 
  84.0 $\pm$ 17.0   &   124.6 $\pm$ 24.6         \\
& [O III]  &    5006.843  &    5001.98   &   -297.0 $\pm$ 21.4   &   63.0 $\pm$ 16.1   &   171.5 $\pm$ 50.6        \\     
& [O I]  &      6300.304 &      6293.58   &   -325.9 $\pm$ 25.3   &   35.4 $\pm$ 7.2   &   251.4 $\pm$ 59.4         \\
& [N II]  &     6548.05  &      6542.02   &   -282.3 $\pm$ 23.0   &   23.0 $\pm$ 
6.2   &   173.3 $\pm$ 54.1       \\
& \Ha  &        6562.8 &        6556.65   &   -287.0 $\pm$ 3.0   &   107.2 $\pm$ 5.3 
  &   125.6 $\pm$ 7.0    \\
& [N II]  &     6583.45  &      6577.3   &   -285.9 $\pm$ 6.9   &   69.1 $\pm$ 6.1   &   160.3 $\pm$ 16.2         \\
& [S II]  &     6716.440 &      6710.21   &   -284.2 $\pm$ 17.1   &   23.1 $\pm$ 5.9   &   136.9 $\pm$ 40.2         \\
& [S II]  &     6730.816  &     6724.6   &   -282.8 $\pm$ 11.0   &   43.3 $\pm$ 6.8   &   142.5 $\pm$ 26.0         \\
& [Fe II] &     7155.1742 &     7148.1   &   -302.4 $\pm$ 21.5   &   14.6 $\pm$ 4.8   &   131.8 $\pm$ 50.4         \\
& [Ca II] &     7291.47  &      7284.61   &   -288.0 $\pm$ 13.9   &   15.6 $\pm$ 
4.3   &   104.4 $\pm$ 34.5       \\
&  [Fe II] &    8616.9498  &    8608.47   &   -301.1 $\pm$ 9.4   &   9.0 $\pm$ 2.6   &   73.4 $\pm$ 30.6  \\
\hline                  
\multirow{11}{*}{\textbf{A4-out}}
& \Hb  &        4861.35 &       4856.45   &   -302.5 $\pm$ 12.1   &   39.4 $\pm$ 6.6   &   142.9 $\pm$ 26.0         \\
& [O III]  &    5006.843  &   5002.35   &   -275.2 $\pm$ 14.4   &   26.2 $\pm$ 5.4   &   145.5 $\pm$ 36.8        \\      
& [O I]  &      6300.304 &      6294.52   &   -281.4 $\pm$ 9.9   &   12.7 $\pm$ 2.0   &   126.8 $\pm$ 23.6         \\
& [N II]  &     6548.05  &      6542.26   &   -271.0 $\pm$ 19.9   &   15.2 $\pm$ 
4.6   &   134.5 $\pm$ 47.3       \\
& \Ha  &        6562.8 &        6556.76   &   -281.8 $\pm$ 2.4   &   104.7 $\pm$ 4.3 
  &   119.0 $\pm$ 5.5    \\
& [N II]  &     6583.45  &      6577.38   &   -282.3 $\pm$ 6.4   &   47.2 $\pm$ 4.6   &   134.6 $\pm$ 15.1         \\
& [S II]  &     6716.440 &      6710.36   &   -277.3 $\pm$ 9.8   &   37.2 $\pm$ 4.9   &   150.8 $\pm$ 23.1         \\
& [S II]  &     6730.816  &     6724.69   &   -278.7 $\pm$ 6.5   &   46.7 $\pm$ 4.6   &   132.9 $\pm$ 15.3         \\
& [Fe II] &     7155.1742 &     7148.51   &   -285.5 $\pm$ 11.7   &   6.2 $\pm$ 1.6   &   99.0 $\pm$ 31.2  \\
& [Ca II] &     7291.47  &      7284.52   &   -292.0 $\pm$ 8.3   &   17.5 $\pm$ 2.1   &   140.4 $\pm$ 19.4         \\
&  [Fe II] &    8616.9498  &    8608.81   &   -289.4 $\pm$ 10.5   &   7.4 $\pm$ 1.5   &   103.5 $\pm$ 24.7         \\

\hline
\end{tabular}
\tablefoot{For atomic lines we could not identify, only observed line properties are provided. The line properties were obtained by fitting a  Gaussian profile. The rest-frame wavelength ($\lambda_{\textrm{rest frame}}$) is obtained from \cite{NIST_ASD}. 
\tablefoottext{a}{We did not correct for the flux loss caused by the low Strehl ratio (25\% at \Ha) because the Strehl ratios are relatively constant for most of line ratios we measured, except for \Ha/\Hb~in A4-out.}
\tablefoottext{b}{The measured FWHM includes instrumental line broadening, which should be considered as the upper limit. The spectral LSF is about 116 km~s$^{-1}$ at \Ha~\citep{Bacon2017}. }
\tablefoottext{c}{The radial velocity of \Hb~at the inner region of A4 is 
not consistent with the rest of the lines. It is due to the residual of stellar \Hb~absorption features that are stronger close to the star. }
}
\end{table*}  


\section{The velocity uncertainty maps}
\begin{figure*}
   \centering
    \includegraphics[width=0.98\textwidth]{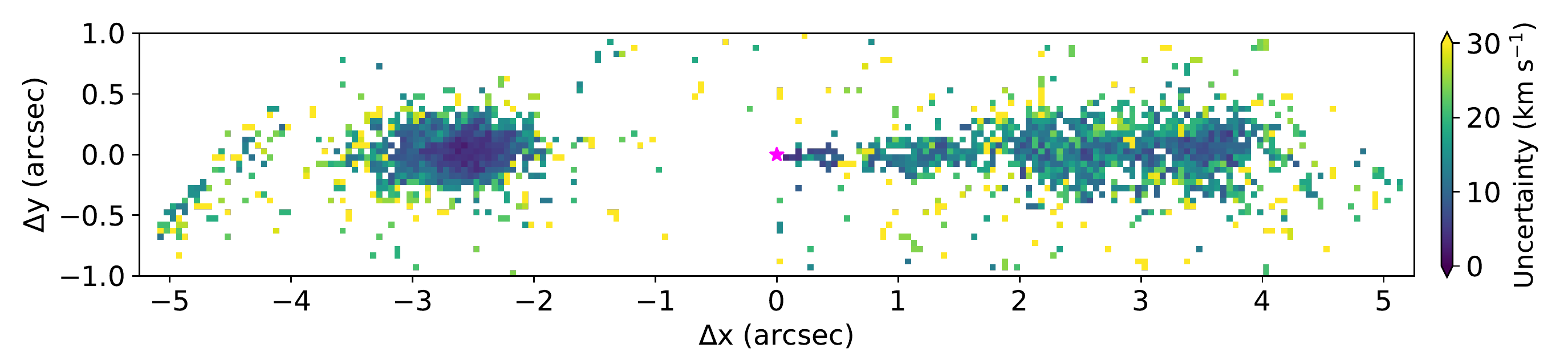}
      \caption{Velocity uncertainty map of the jet observed by MUSE, traced by [S II] $\lambda$630 ($\Delta$x~\textless~0 ) and \Ha~($\Delta$x~\textgreater~0).
      }
         \label{velocity_uncertainty_map}
\end{figure*}

Figure~\ref{velocity_uncertainty_map} shows the uncertainty in the radial 
velocity. The procedure to construct the radial velocity map and its uncertainty map is explained in Sect.~\ref{sec:velocity_map}. We note that only the uncertainty caused by the measurement noise is shown. We cannot estimate the residual offset after the extra wavelength calibration due to 
the lack of prominent emission or absorption features (see also, Sect.~\ref{subsec:velocity_unvertainty}).

\section{Physical conditions estimation}
\label{appsec:phy_condition}

The physical conditions were estimated based on \cite{Hartigan1994}, except for the observed electron density ($n_{\rm e}$) in the post shock \citep{Proxauf2014}. The derived physical parameters are summarized in Table~\ref{tab:ratio_and_conditions}. As shown in the Fig.3 in \cite{Ellerbroek2014}, we estimated the pre-shock density ($n_{\rm H,~pre}$) and the magnetic field strength $B$ based on the line ratios we measured. The shock velocities ($v_{\rm shock}$) listed in Table~\ref{tab:ratio_and_conditions} were suggested by the line ratios of [N~II] $\lambda$658 and [O~I] 
$\lambda$630, which agreed with the predictions of best represented models listed in Table~\ref{tab:ratio_and_conditions}. 

We estimated the ionization fraction $<I>$ based on the line ratios of [N~II] $\lambda$658 and [O~I] $\lambda$630 (see, Fig.~6 in \cite{Hartigan1994}), using $n_{\rm H,~pre}$ listed in Table~\ref{tab:ratio_and_conditions}. We adopted an average magnetic field strength of 30 $\mu$G if $B$ ranges from 10 to 100~$\mu$G. The compression factor can then be estimated from the shock velocity (see, Fig.~17 in \cite{Hartigan1994}). The average density ($<n_{\rm H}>$) can be expressed as $<n_{\rm H}> = n_{e}<C>^{-1/2}<I>^{-1}$ \citep{Hartigan1994}.


\end{appendix}

\end{document}